\documentclass[eqsecnum,amsmath,preprintnumbers,superscriptaddress,nofootinbib,aps,11pt]
{revtex4}
\pdfoutput=1
\usepackage{tikz}
\usetikzlibrary{calc,trees,positioning,arrows,chains,shapes.geometric,decorations.pathreplacing,decorations.pathmorphing,shapes,matrix,shapes.symbols}

\usepackage[T1]{fontenc}
\usepackage[utf8]{inputenc}
\usepackage{bm}
\usepackage{MnSymbol}
\usepackage{amsmath}
\usepackage{graphicx}
\usepackage{feynmp}
\usepackage{hyperref}
\usepackage{bbold}
\usepackage{eufrak}
\usepackage{upgreek}
\usepackage{stmaryrd,scalerel}
\usepackage{mathtools}
\usepackage{enumitem}
\usepackage{color} 
\usepackage{etex}
\usepackage{morefloats}

 \usepackage{longtable}

\def\e{{\rm{e}}}
\def\ii{{\rm{i}}}
\def\d{{\rm{d}}}

\def\beq{\begin{equation}}
\def\eeq{\end{equation}}
\def\bea{\arraycolsep .1em \begin{eqnarray}}
\def\eea{\end{eqnarray}}
\def\Tr{{\rm Tr}}
\def\tr{{\rm tr}}

\def\!!!{\stackrel{!}{=}}

\def\a{\alpha}
\def\b{\beta}

\def\STr{ {\rm STr } }

\def\G{ \mathcal{G}}

\def\nn{ \nonumber \\}

\def\eq#1{(\ref{#1})}

\def\s0#1#2{\mbox{\small{$ \frac{#1}{#2} $}}}
\def\0#1#2{\frac{#1}{#2}}

\def\grgl{\:\hbox to -0.2pt{\lower2.5pt\hbox{$\sim$}\hss}{\raise3pt\hbox{$>$}}\:}
\def\klgl{\:\hbox to -0.2pt{\lower2.5pt\hbox{$\sim$}\hss}{\raise3pt\hbox{$<$}}\:}

\def\Reg{\mathcal{R}}
\def\reg{R}

\begin{document}
\title{Essential Quantum Einstein Gravity}
\author{Alessio Baldazzi}
\email{abaldazz@sissa.it}
\affiliation{SISSA -- International School for Advanced Studies \& INFN, via Bonomea 265, I-34136 Trieste, Italy}
\author{Kevin Falls}
\email{kfalls@sissa.it}
\affiliation{SISSA -- International School for Advanced Studies \& INFN, via Bonomea 265, I-34136 Trieste, Italy}
\date{\today}
\begin{abstract}
The non-perturbative renormalisation of quantum gravity is investigated allowing for the metric to be reparameterised along the RG flow, such that only the essential couplings constants are renormalised. This allows us to identify a universality class of quantum gravity  which is guaranteed to be unitary, since the physical degrees of freedom are those of general relativity without matter and with a vanishing cosmological constant.
Considering all diffeomorphism invariant operators with up to four derivatives, only Newton's constant is essential at the Gaussian infrared fixed point associated to the linearised Einstein--Hilbert action. The other inessential couplings can then be fixed to the values they take at the Gaussian fixed point along the RG flow within this universality class. 
In the ultraviolet, the corresponding beta function for Newton's constant vanishes at the interacting Reuter fixed point. The properties of the Reuter fixed point are stable between the Einstein--Hilbert approximation and the approximation  including all diffeomorphism invariant four derivative terms in the flow equation.
Our results suggest that Newton's constant is the only relevant essential coupling at the Reuter fixed point.  Therefore, we conjecture that quantum Einstein gravity, the ultraviolet completion of Einstein's theory of general relativity in the asymptotic safety scenario, has no free parameters in the absence of matter and in particular predicts a vanishing cosmological constant. 
\end{abstract}
\maketitle
%
\section{Introduction}
\label{sec:intro}

Wilson's exact renormalisation group (RG)~\cite{Wilson:1973jj} provides a framework to construct a consistent quantum field theory (QFT) that describes gravity. This possibility, known as asymptotic safety, relies on the gravitational couplings exhibiting an ultraviolet (UV) fixed point that allows the UV cut-off to be removed while keeping physical quantities finite~\cite{Weinberg}. The theory can then be defined as a perturbation of the fixed point along a renormalisable trajectory that leaves the UV fixed point and evolves towards the infrared (IR), where it is identified with the renormalised theory. In this framework~\cite{perbook,rsbook}, the number of free dimensionless parameters is one fewer than the number of relevant couplings at the fixed point, which parameterise the UV critical surface formed from all renormalisable trajectories.

So far, the evidence suggests that there is such a fixed point, known as the Reuter fixed point~\cite{Reuter:1996cp,Souma:1999at,Reuter:2001ag,Lauscher:2001ya,Litim:2003vp},  and that it possesses three relevant couplings in pure gravity~\cite{lauscher,Machado:2007ea,CPR1,Codello:2008vh,bms,bms2,Falls:2013bv,Falls:2014tra,Denz:2016qks,FKL2018,Falls:2018ylp,Falls:2020qhj}. However, not all couplings need to reach a fixed point for the theory to be asymptotically safe, since one has the freedom to perform field reparameterisations which can be used to eliminate the so-called inessential couplings  from the RG equations~\cite{Weinberg}.
 The inessential couplings do not appear in expressions for observables, such as cross sections and reaction rates, and, therefore, can take different values without affecting the physics.
 Couplings, therefore, fall into two classes: the essential couplings $\lambda_a$ which enter into expressions for observables and the inessential couplings $\zeta_\a$ which are scheme dependent and unphysical.
 Consequently, the scaling behaviour of inessential couplings is entirely scheme dependent and they must not be included in the set of relevant couplings~\cite{Wegner1974}. It follows that a coupling that may appear relevant could turn out to be inessential and,  therefore, does not contribute to the counting of free parameters.
 Although the potential existence of inessential couplings has been pointed out~\cite{perbook,rsbook,Percacci:2004sb,Dietz:2013sba,Steinwachs:2020jkj}, they have been almost universally ignored in investigations of asymptotic safety. In particular, attempts to find a suitable fixed point have required fixed points for all gravitational couplings, included in a given approximation, instead of incorporating field reparameterisations into the RG equations and checking which of the couplings are inessential. Here we shall remedy this oversight by incorporating field reparameterisations in the gravitational RG equations which allow us to eliminate the inessential couplings from the flow equations.  To do so we will utilise the essential RG approach, which has been put forward in~\cite{Baldazzi:2021ydj}, where we only compute the flow of the essential couplings.

Our strategy will be to adapt the minimal essential scheme devised in~\cite{Baldazzi:2021ydj}, in the context of scalar field theories, to remove all inessential couplings in pure gravity. This can be carried out order by order in the derivative expansion, where only terms with up to $s$-derivatives of the fields are included in the effective action.
At each order $s$ the minimal essential scheme is implemented by identifying the inessential couplings at a  Gaussian fixed point of the theory and fixing their values, such that one obtains beta functions for the essential couplings only. An important point is that this scheme involves a specification of the kinematical degrees of freedom, since it assumes that the degrees of freedom are those of the Gaussian fixed point. This implies that the minimal essential scheme can break down a finite distance from the Gaussian fixed point and, thus, cannot describe all possible non-perturbative behaviour. However, one can then instead identify inessential couplings at other points in theory space, which, while technically more involved, would allow the essential RG to describe all regions of theory~space.

   For a scalar field there are Gaussian fixed points associated to kinetic operators $(-\partial^2)^{s/2}$ for every even integer $s$, which involve different degrees of freedom.
As such, there is a minimal essential scheme associated to each Gaussian fixed point, that is physically distinct since they are associated with different degrees of freedom. Within a given minimal essential scheme, the RG flow is then constrained to the physical theory space associated to those degrees of freedom. In other words, the minimal essential scheme restricts the RG flow to a universality class that contains the corresponding Gaussian fixed point.  Although, typically, RG studies are concerned with the universality class involving the Gaussian fixed point for which $s=2$, one can also study universality classes associated to higher derivative theories~\cite{Safari:2017irw,Safari:2017tgs}. When one utilises the minimal essential scheme for $s=2$, the Gaussian fixed points for higher derivative theories are excluded. Therefore, this choice is not without physical consequences since by adopting a minimal essential scheme we focus our attention on possible fixed points in a specific universality class rather than attempting to find all possible fixed points.  

For quantum gravity, we will consider the universality class of quantum Einstein gravity (QEG) meaning that it is associated to the quantisation of the physical degrees of freedom associated to Einstein's theory of gravity.  As such, in this paper by the Gaussian fixed point (GFP) in the context of gravity we refer to the one associated to the linearised Einstein--Hilbert action unless otherwise stated. Here we should point out that we mean something more specific (but perhaps more deserving of the name) by QEG than the more broad definition given, e.g., in~\cite{Reuter:2012id}.  In particular, we do not only specify the fields and symmetries, in terms of which we parameterise the theory, but also the physical degrees of freedom. For example, a quantisation of higher derivative gravity~\cite{Stelle:1976gc} can be carried out by quantising the metric assuming diffeomorphism invariance, but it is a quantisation of more degrees of freedom than Einstein's theory. This shift of emphasis to the physical degrees of freedom and the physical essential couplings  will bring our investigation of asymptotic safety closer to the original formulation~\cite{Weinberg} by S.~Weinberg: a move which has been  strongly encouraged recently~\cite{Donoghue:2019clr}.

To set the stage, in the Section~\ref{sec:reviewscacase}  we give a short review of the essential RG technique, which generalises the usual approach to the exact (aka the non-perturbative functional) RG (see~\cite{Morris:1998da,Berges:2000ew,Bagnuls:2000ae,Pawlowski:2005xe,Rosten:2010vm,Delamotte:2007pf,Dupuis:2020fhh} for reviews) for the effective average action (EAA) by allowing for field to be reparameterised along the RG flow.  In Section~\ref{sec:weinberg} we revisit S.~Weinberg's formulation of asymptotic safety which emphasises the manner in which essential couplings enter expressions for observables.
In Section~\ref{sec:genflowquagra} we derive the generalised flow equation for quantum gravity that takes into account the freedom to reparameterise the quantum metric along the RG flow.
Indeed, the flow equation will contain a new ingredient: the RG kernel. This quantity encodes the description about how the fields are reparameterised along the flow.
Then, we write down a systematic derivative expansion of the diffeomorphism invariant part of the EAA and the covariant RG kernel. In particular, we expand the EAA to fourth order in derivatives and the RG kernel to second order. 
In Section~\ref{sec:vaceniness}, we analyse the GFP properties: in particular, from this analysis we determine that the vacuum energy is inessential. The advantage of studying the GFP consists of the fact that it is a free fixed point and the results can be obtained without relying on approximations.
After having found the inessential couplings at the GFP, in Section~\ref{sec:minessschemequagra} we discuss the properties of the universality class that contains the GFP and all the trajectories that have the same essential couplings. In such a subspace of the theory space, the propagator evaluated on conformally flat spacetime possesses the same form as  the one at the GFP. 
Up to order four in the derivative expansion (apart from the topological Gauss--Bonnet term) only Newton's constant is essential in this universality class. 
In particular, any fixed point on these trajectories has the degrees of freedom of the GFP.
In Section~\ref{sec:reuterfixpoint}, we study the RG flow of QEG in the minimal essential scheme at orders two and four of the derivative expansion. Our investigation confirms the existence of the Reuter fixed point: this implies that higher derivative terms coming from the operators $\sqrt{\det g}\, R^2$ and $\sqrt{\det g}\, R_{\mu\nu}R^{\mu\nu}$ are inessential in the universality class of the GFP. Moreover, this means that, contrary to the expectations based on  perturbative renormalisability, the existence of the Reuter fixed point in the minimal essential scheme suggests that a possible UV-completion of the gravitational theory does not require additional degrees of freedom. 
In Section~\ref{sec:conclusion} we draw our conclusions and discuss the outlook for future investigations of quantum gravity using the essential RG. The derivation of RG equations in the minimal essential scheme for QEG at fourth order in the derivative expansion for a generic dimension and a generic regulator cut-off  are presented in Appendix~\ref{app:gravitytrace}.

\section{Essentials of the Essential Renormalisation Group }
\label{sec:reviewscacase}
In this section we review the essential RG approach~\cite{Baldazzi:2021ydj} using the case of a single scalar field to avoid overloading notation and technicalities.
The generalisation for gravity will be developed in the rest of the paper starting in Section~\ref{sec:genflowquagra}.

Ultimately in QFT we are interested in expectation values of observables
\begin{equation} \label{Expectation_O}
\mathcal{O} = \langle \hat{\mathcal{O}} \rangle :=  \mathcal{N}   \int ({\rm{d}} \hat{\chi}) ~\hat{\mathcal{O}}[\hat{\chi}]~ {\rm e}^{-S[\hat{\chi}]}  \,,
\end{equation}
where $\mathcal{N}^{-1} =  \int ({\rm{d}} \hat{\chi}) \, {\rm e}^{-S[\hat{\chi}]}$ and $\hat{\mathcal{O}}[\hat{\chi}] = \hat{\mathcal{O}}$ is an observable expressed as functional of the fields $\hat{\chi}$.
The essential RG is a method to eventually compute~\eqref{Expectation_O} that makes use of the generalised exact RG equation for the EAA, which depends on the RG scale $k$.
The EAA obtains a dependence on the RG scale $k$ from two sources.  First, the EAA depends on $k$ due to the presence of a momentum-dependent IR cut-off
 \begin{align}
 \Reg_k(x_1,x_2) &=  k^2 \, \reg(\Delta/k^2) \delta(x_1,x_2)\nonumber \\
 & =  k^2 \int \frac{\d^dp}{(2\pi)^d} \, \, \reg(p^2/k^2 ) \, \e^{\ii p_\mu(x^\mu_1 - x^{\mu}_2)}\,,
 \end{align}
which  implements the coarse-graining procedure, cutting off low momentum modes in the functional integral~\eqref{eq:defEAA} that defines the EAA.
This is achieved by choosing the dimensionless function $R(p^2/k^2)$ such that it vanishes in the limit $p^2/k^2 \to \infty$, while for $p^2/k^2 \to 0$ it has a non-zero limit $R(0)>0$, ensuring the suppression of IR modes. This realises Wilson's picture of the RG which integrates out UV modes successively as $k$ is lowered. The second source of $k$ dependence comes
from the liberty to perform field reparameterisations along the flow parameterised by a $k$-dependent diffeomorphism  $\hat{\phi}_k[\hat{\chi}]$  of configuration space which we integrate over in~\eqref{Expectation_O}.   This is achieved by considering a generating functional for correlation functions of the $k$-dependent fields $\hat{\phi}_k[\hat{\chi}]$ rather than the $k$-independent fields $\hat{\chi}$.
Explicitly this functional is the generalised EAA action $\Gamma_k[\phi]$ defined by the functional integro-differential equation
\bea \label{eq:defEAA}
\e^{-\Gamma_k[\phi]} &:=&
  \int ({\rm{d}} \hat{\chi}) ~ {\rm e}^{-S[\hat{\chi}] + (\hat{\phi}_k[\hat{\chi}] - \phi) \cdot \frac{\delta}{\delta \phi} \Gamma_k[\phi]
 - \frac{1}{2}  ( \hat{\phi}_k[\hat{\chi}] - \phi) \cdot  \Reg_k \cdot (\hat{\phi}_k[\hat{\chi}] - \phi) }\,,
 \eea
from which it follows that
\beq
\phi = \langle \hat{\phi}_k \rangle_{\phi,k} \,,
\eeq
where
\beq \label{eq:ObsJR}
 \langle \hat{\mathcal{O}} \rangle_{\phi,k} :=   \e^{\Gamma_k[\phi]} \int ({\rm{d}} \hat{\chi}) ~ {\rm e}^{-S[\hat{\chi}] + (\hat{\phi}_k[\hat{\chi}] - \phi) \cdot \frac{\delta}{\delta \phi} \Gamma_k[\phi]
 - \frac{1}{2}  ( \hat{\phi}_k[\hat{\chi}] - \phi) \cdot  \Reg_k \cdot (\hat{\phi}_k[\hat{\chi}] - \phi) }  \hat{\mathcal{O}} [\hat{\chi}]
\eeq
is the $\phi$ and $k$ dependent expectation value.
In the limit $k\to0$ the cut-off vanishes and EAA reduces to the one-part irreducible effective action $\Gamma[\phi]= \Gamma_0[\phi]$ for the field $\hat{\phi}_0$.
In the opposing limit $k \to \infty$ the EAA reduces to the bare action written in terms of the fields $\hat{\phi}_\infty$.
  Let us note that we could additionally make a change of integration variables in the RHS of~\eqref{eq:defEAA} which keeps $\Gamma_k[\phi]$ invariant provided we make this change everywhere including in the measure. Here we are keeping the integration variables $\hat{\chi}$ and the bare action $S[\hat{\chi}]$  $k$-independent, such that the $k$-dependence comes only from the regulator and the composite fields $\hat{\phi}_k[\hat{\chi}]$. Since we are ultimately interested in computing observables~\eqref{eq:ObsJR} at vanishing regulator and on the equations of motion for $\Gamma_0[\phi]$ we will recover~\eqref{Expectation_O} independently of the regulator and the parameterisation  $\hat{\phi}_k[\hat{\chi}]$. For more details we refer the reader to~\cite{Baldazzi:2021ydj}. 

 In general, $\Gamma_k[\phi]$ will depend on all couplings compatible with the symmetries of the theory. Incorporating   $k$-dependent field reparameterisations into the EAA formalism has first been considered in~\cite{Gies:2001nw} to describe bound states.
In the essential RG, the utility of $\hat{\phi}_k[\hat{\chi}]$ is that we may choose to reparameterise the field to fix the values of inessential couplings which, by definition, are simply those couplings that depend on the form of $\hat{\phi}_k[\hat{\chi}]$. Since observables~\eqref{Expectation_O} are invariant under a change in $\hat{\phi}_k[\hat{\chi}]$, they do not depend on the inessential couplings.
Any scheme which fixes or otherwise specifies the flow of all inessential couplings is an essential scheme. Thus, in an essential scheme we only compute the flow of essential couplings $\lambda_a(k)$, i.e., those which ultimately enter into observables~\eqref{Expectation_O}.

The generalised flow equation satisfied by $\Gamma_k[\phi]$ is given by~\cite{Pawlowski:2005xe}
\begin{equation} \label{scadimensionfull_flow}
\left( \partial_t  +   \Psi_k[\phi] \cdot  \frac{\delta }{\delta \phi}  \right) \Gamma_k[\phi]
= \frac{1}{2} \Tr \,  \mathcal{G}_k[\phi]  \left( \partial_t  + 2  \cdot \frac{\delta}{\delta \phi}  \Psi_k[\phi] \right)\cdot \mathcal{R}_k \,,
\end{equation}
where $t := \log(k/k_0)$, with $k_0$ some physical reference scale, under the trace appearing in the RHS
\beq
\mathcal{G}_k[\phi] :=  (\Gamma^{(2)}_k[\phi] + \mathcal{R}_k)^{-1}\,
\eeq
 is the IR regularised propagator, with $\Gamma^{(2)}_k[\phi] $ denoting the hessian of the EAA with respect to the field $\phi(x)$,
and
\begin{equation}
\Psi_k[\phi]  :=   \langle \partial_t \hat{\phi}_k[\hat{\chi}] \rangle_{\phi,k}\,
\end{equation}
is the RG kernel which takes into account the $k$-dependent field reparameterisations.
The flow Equation~\eqref{scadimensionfull_flow}  reduces to the standard flow for the EAA~\cite{Wetterich:1992yh,Morris:1993qb} when $\Psi_k =0$ and can be understood as the counterpart to the generalised flow for the Wilsonian effective action~\cite{Wegner1974}.
   
By choosing $\Psi_k[\phi]$ we can specify the flow of inessential couplings $\zeta$, which by their definition are those for which~\cite{Baldazzi:2021ydj}
\begin{equation} \label{Redundant_operator}
\zeta \frac{\partial}{\partial \zeta} \Gamma_k[\phi]  = \mathcal{T}[\Phi] :=  \Phi[\phi] \cdot  \frac{\delta }{\delta \phi}   \Gamma_k[\phi] - \Tr \, \mathcal{G}_k[\phi] \cdot \frac{\delta}{\delta \phi}   \Phi[\phi]  \cdot \mathcal{R}_k \,,
\end{equation}
for some quasi-local field $\Phi[\phi]$.   Equation~\eqref{Redundant_operator} follows from the definition of an inessential coupling  and, therefore, allows them to be identified. The operator appearing on the RHS of~\eqref{Redundant_operator} is the redundant operator conjugate to the inessential coupling $\zeta$. The first term in the RHS of~\eqref{Redundant_operator} is tree-level and is simply proportional to the equation of motion for $\Gamma_k[\phi]$ and survives in the limit $k\to 0$. The second term instead vanishes as when the regulator vanishes as $k \to 0$. 
Within perturbation theory, in the vicinity of a Gaussian fixed point the second term will be sub-leading, since it is loop correction being proportional to Planck's constant $\hbar$.   In general, there will be an inessential coupling associated to every linearly independent  quasi-local field $\Phi_k[\phi]$ which generates an independent field reparameterisation. Although the possible field reparameterisations $\Phi_k[\phi]$ are themselves independent of the position in theory space, it is important to stress that the redundant operator depends on the EAA $\Gamma_k[\phi]$ and, thus,
the identification of inessential couplings will depend on the form of the EAA. Thus, couplings which may be inessential at one fixed point can be essential at others. As an example at the GFP
\beq \label{Scalar_GFP}
 \Gamma_k^{\rm GFP}[\phi] = \frac{\zeta}{2} \int \d^dx \,\phi (- \partial^2) \phi\,,
\eeq
the coefficient of $\zeta$ of the kinetic term is inessential. This can be understood since  on the equations of motion $\partial^2 \phi = 0$ the kinetic term vanishes. Changing the value of $\zeta$ corresponds to moving along a line of equivalent fixed points.
However, if we consider the fourth order GFP 
\beq \label{Scalar_GFP_4}
 \Gamma_k^{\rm GFP4}[\phi] = \frac{\zeta}{2} \int \d^dx \, \phi (- \partial^2)^2 \phi \,,
 \eeq
 the operator $ \frac{1}{2} \int \d^dx \,\phi (- \partial^2) \phi$ is not redundant since it does not vanish on the equations of motion $(\partial^2)^2 \phi=0$ for~\eqref{Scalar_GFP_4}. Here we also see the connection between inessential couplings and the number of degrees of freedom. For the 
fourth order theory we have two propagating degrees of freedom which are massless at the fixed point~\eqref{Scalar_GFP_4}.  By adding the term with two derivatives, the action becomes
\beq \label{Scalar_GFP_4_mass}
 \Gamma_k[\phi] = \frac{\zeta}{2} \int \d^dx \, \phi (- \partial^2)( - \partial^2  + m^2 )\phi \,,
 \eeq
where $m^2$ is an essential coupling being identified as a mass for one of the degrees of freedom. Let us also note that at the GFP~\eqref{Scalar_GFP} the higher order term $\int \d^dx \, \phi (- \partial^2)^2 \phi $ is redundant since it vanishes on the equations of motion $\partial^2 \phi = 0$. This reflects the fact that by starting with only one propagating degree of freedom we cannot gain more degrees of freedom along the RG flow.

 Since the terms involving $\Psi_k[\phi]$ in~\eqref{scadimensionfull_flow} have the form of a redundant operator, the liberty to choose $\Psi_k[\phi]$ is precisely the liberty to specify the flow of all inessential couplings. Thus, for each inessential coupling we specify an RG condition,  understood as a constraint  on the form of $\Gamma_k[\phi]$ along the RG flow, then we solve the flow equation under this condition for the beta functions of the essential couplings and gamma functions which parametrise $\Psi_k[\phi]$. Different essential schemes correspond to different sets of RG conditions for the inessential couplings.
From a geometric point of view, we can think of reparameterisations as local frame transformations on configuration space that are analogous to gauge transformations~\cite{Latorre:2000qc,Baldazzi:2021ydj}. RG conditions are, therefore, analogous to gauge fixing conditions which fix a particular frame, as with gauge conditions we typically want to find RG conditions that minimise the complexity of a given observable.

Since the form of the redundant operators~\eqref{Redundant_operator} depend on $\Gamma_k[\phi]$ in practice the simplest scheme to implement is the minimal essential scheme which sets all inessential couplings at the GFP~\eqref{Scalar_GFP}
to zero (apart from the coefficient of the kinetic term which is canonically normalised). One can show that this is achieved by setting all terms in the $\Gamma_k[\phi]$, which can be brought into the form $ \int \d^dx \,\Phi \,\partial^2\phi$ by an integration by parts, to zero. In other words, in the minimal essential scheme we put to zero any term in $\Gamma_k[\phi]$ that vanishes when we apply the equations of motion at the GFP apart from the canonically normalised kinetic term~\eqref{Scalar_GFP} itself.
Thus, while at order $\partial^2$ in a derivative expansion, $\Gamma_k$ can assume the form
\begin{align}
\Gamma_k = \int \d^dx \, \left\{ V_k(\phi) + \frac{1}{2}z_k(\phi) \left( \partial_\mu \phi \, \partial_\mu \phi \right) \right\}\, + O(\partial^4),
\end{align}
in the minimal essential scheme, the EAA reduces to
\begin{align}\label{eq:EAAorder4}
\Gamma_k &= \int \d^dx \, \left\{ V_k(\phi) + \frac{1}{2} \left( \partial_\mu \phi \, \partial_\mu \phi \right) + O(\partial^4)
\right\} \,,
\end{align}
which involves only one function, namely the effective potential $V_k(\phi)$.
In order to solve the flow equation up to order $\partial^2$ in the essential RG, the RG kernel must have the form
\begin{align}
\Psi_k(x) =  F_k(\phi) + O(\partial^2) \,.
\end{align}
Although typically we would have a non-linear dependence on $z_k(\phi)$ in the flow equation coming from the propagator $\mathcal{G}_k$, in the minimal essential scheme this dependence is absent.
Thus, by adopting the minimal essential scheme we trade a non-linear dependence on $z_k(\phi)$ in the flow equation for a linear dependence of $F_k(\phi)$. More generally, in the minimal essential scheme the  $\mathcal{G}_k[\phi]$ evaluated at any constant value of the field $\phi(x) = \bar{\phi}$ has the~form
\beq \label{Minimal_Propagator_Scalar}
\mathcal{G}_k[\bar{\phi}] = \frac{1}{ -\partial^2 + \mathcal{R}_k   + V''_k(\bar{\phi})} \,,
\eeq
where $V''_k(\bar{\phi})$ is the second derivative of the potential.

The simplified form of the propagator~\eqref{Minimal_Propagator_Scalar}, which continues to hold at any order of the derivative expansion, produces simplifications in practical calculations, and maintains a form that manifestly contains only physical degrees of freedom which are present at the GFP~\eqref{Scalar_GFP}. This implies, for example, the absence of ghosts and tachyons, and it constraints our theory to stay in the subspace of the theory space where the degrees of freedom are the same of the GFP. As we will see, these features can also be guaranteed for the graviton propagator.  What cannot be guaranteed is that there also exists other fixed points apart from the GFP in this subspace. Thus, by adopting the minimal essential scheme, we limit our search for additional fixed points by constraining the propagating degrees of freedom.
\section{Weinberg's Formulation of Asymptotic Safety}
\label{sec:weinberg}
Having reviewed the essential RG, let us now discuss
the criteria of asymptotic safety as formulated by Weinberg~\cite{Weinberg} and how it is realised by solving the flow equation for the EAA within an essential scheme. The criteria necessitate that we have a UV-complete QFT where there is no UV cut-off, which is achieved if the theory lies on an RG trajectory that originates from a UV fixed point. However, as has been emphasised recently~\cite{Donoghue:2019clr}, Weinberg's formulation is more precise since it is concentrated on the absence of unphysical UV divergences in physical quantities, such as reaction rates, rather than on the behaviour of correlation functions of fields $\hat{\phi}$. This is important since correlation functions depend on inessential couplings $\zeta_\alpha$. In a scheme where we do not specify the values of inessential couplings but compute their flow, we are at the very least making our life harder unnecessarily. In the worst-case, an inessential coupling may not reach a fixed point and thus in such a scheme asymptotic safety could be obscured. In an essential scheme, we only compute the flow of the essential couplings and, thus, avoid these matters.

The divergences, which are absent in asymptotic safety, are those we expect to appear if we only have an effective theory that involves an artificial UV cut-off $\Lambda_{\rm UV}$, characterising our ignorance of physics on small distances $\ell < 1/\Lambda_{\rm UV}$. An effective theory will break down as energies approach the cut-off scale and we will, therefore, encounter unphysical divergences.
In an asymptotically safe theory, such divergences should be absent since we have sent $\Lambda_{\rm UV} \to \infty$.
Indeed, the form of the flow Equation~\eqref{scadimensionfull_flow} assumes that the limit $\Lambda_{\rm UV} \to \infty$ has been taken and would take a modified form if an independent UV cut-off were introduced~\cite{Morris:1993qb,Morris:2015oca}.
 Asymptotic safety requires that  as we take some characteristic energy scale $E \to \infty$ observables (such as reaction rates) scale as
\beq
\mathcal{O} \sim E^{D} \,,
\eeq
where $D$ is the dimension of $\mathcal{O}$.  This means in particular that dimensionless quantities will not diverge even when we take $E \to \infty$ and, thus, at high energies the theory is scale invariant.
Note that asymptotic safety is a rather  generic requirement that we impose to be ``reasonably sure'' that there are no divergences in physical quantities related to the theory breaking down at a finite energy scale. On the one hand, asymptotic safety does not rule out {\it all} divergent behaviour, since unobservable correlation functions can diverge at finite energies even if the theory is well defined at all energies.
On the other hand, asymptotic safety does {\it not} guarantee that a theory is physically acceptable since, for example, there can be asymptotically safe theories that are not unitary~\cite{Rosten:2011mf}, the simple example being a free theory with four derivatives.

If we were handed the full quantum effective action $\Gamma$ and computed observables from it directly, the coupling constants entering the expression for an observable would be the essential couplings $\lambda_a^{\rm phys.} \equiv \lambda_a(0)$ evaluated at $k=0$. One may then wonder what the link is to a fixed point of the exact RG obtained in the opposing limit $k \to \infty$. In particular one may worry that observables can depend on additional energy scales $E_n$ in addition to the scale $E$ which we take to infinity.
To understand the connection, note that if we supply an initial condition for the flow at a scale $k=\mu$, the flow equation supplies a function
\beq
\lambda_a^{\rm phys.} = \lambda_a^{\rm phys.}(\lambda_b(\mu) , \mu) \,,
\eeq
since by integrating the flow for a given initial condition we will obtain $\lambda_a$ when we arrive to $k=0$.
Therefore, we can write any observable which depends on energy scales $E$ and $\{E_n\}$ and the physical couplings $\lambda_a^{\rm phys.}$   as a function
\beq
\mathcal{O} = \mathcal{O}(E,\lambda_a(\mu) , \mu, E_n) \,,
\label{eq:observable}
\eeq
where $ \mathcal{O}(E,\lambda_a(\mu) , \mu, E_n) $ is independent of $\mu$ by construction meaning.
On the other hand, dimensional analysis means that we can also write
\beq \label{Omu}
\mathcal{O} = \mu^D \tilde{\mathcal{O}}(E/\mu,\tilde{\lambda}_a(\mu),E_n/\mu) \,,
\eeq
where $\tilde{\lambda}_a(\mu) = \mu^{-d_a} \lambda_a(\mu)$ are the dimensionless couplings and $d_a$ is the mass dimension of the coupling $\lambda_a$.
Generically, the functions for the dimensionless observables  $ \tilde{\mathcal{O}}$ will be finite for finite values of its arguments, while if one argument were to diverge then generically we expect  $\tilde{\mathcal{O}}$ to become singular.
Now, since $\mathcal{O}$ is independent of $\mu$, we can set $\mu = E$, such that
\beq
\mathcal{O} = E^D \tilde{\mathcal{O}}(1,\tilde{\lambda}_a(E),E_n/E) \,.
\eeq
Then, it is clear that the limit $E\to \infty$ only exists if the limit $\lim_{\mu \to \infty} \tilde{\lambda}_a(\mu)$ exists.
If a subset of the dimensionless essential couplings $\tilde{\lambda}_a(E)$ diverges at some finite $E = \Lambda_{\rm UV}$, then we expect the observable to be singular at this point.  However, if all the couplings $\tilde{\lambda}_a(\mu)$ remain finite for $\mu \to \infty$, such that they reach a UV fixed point $\lim_{\mu \to \infty}  \tilde{\lambda}_a(\mu) =  \lambda_a^\star$, then
\beq
\lim_{E \to \infty} \mathcal{O} = E^D \tilde{\mathcal{O}}(1,\tilde{\lambda}_a^\star,0) \,,
\eeq
which is exactly the requirement of asymptotic safety.
An important point is that, since the RHS of~\eqref{Omu} is independent of $\mu$, if we would send $E_n$ to infinity instead of $E$ we could then identify $\mu = E_n$ and reach the conclusion that  $\mathcal{O} \sim E_n^D$ as $E_n \to \infty$.

Crucially, it is only the essential couplings that need to attain a UV fixed point. Indeed, inessential couplings $\zeta_\alpha$ are simply not present in physical observables~\eqref{eq:observable} and, therefore, their behaviour is not restricted a priori. All of these remarks apply to asymptotically safe theories in general, in the remainder of this paper we will develop the formalism to  investigate asymptotic safety in quantum gravity within an essential scheme.
\section{Generalised Flow Equation and Essential Schemes for Quantum Gravity}
\label{sec:genflowquagra}
In this section, we will derive the generalised flow equation for quantum gravity from which we use to apply the essential RG method in order to investigate asymptotic safety. This construction generalises the formalism introduced in~\cite{Reuter:1996cp} by allowing for the field redefinitions at the heart of the essential RG.
For quantum gravity the EAA is denoted $\Gamma_k[f;\bar{g}]$, where  $f= \{ g_{\mu\nu}, c^{\mu}, \bar{c}_{\mu} \}$ denotes the set of mean fields, $g_{\mu\nu}$ is the (mean) metric, and $ c^{\mu}$ and  $\bar{c}_{\mu}$ are the (mean) anti-commuting ghost and anti-ghost. In addition to the mean fields, $\Gamma_k[f;\bar{g}]$ also depends on an auxiliary background metric $\bar{g}_{\mu\nu}$ in order to conserve background covariance. The EAA for gravity is defined analogously to the case of the scalar field~\eqref{eq:defEAA} through the functional integral
\beq \label{EAA_def}
{\rm{e}}^{- \Gamma_k[f,\bar{g}]} = \int \left({\rm{d}} \hat{\chi}\right)\, {\rm{e}}^{-S[\hat{\chi}; \bar{g} ]} {\rm{e}}^{(\hat{f}_k[\hat{\chi}] - f)\cdot \frac{\delta }{\delta f}  \Gamma_k[f;\bar{g}] }
{\rm{e}}^{- \frac{1}{2} (\hat{f}_k[\hat{\chi}] - f)\cdot  \mathcal{R}_k[\bar{g}] \cdot  (\hat{f}_k[\hat{\chi}] - f) } \,,
\eeq
where $\hat{\chi}$ are a set of fields which parameterise the fields $\hat{f}_k[\hat{\chi}]=  \{ \hat{g}_{\mu\nu \, k}[\hat{\chi}], \hat{c}^{\mu}_{\,k}[\hat{\chi}], \hat{\bar{c}}_{\mu \, k}[\hat{\chi}] \}$, such that the latter defines a $k$-dependent diffeomorphism of the configuration space to itself. Formally, since the configuration space involves the ghost fields, it is a super-manifold. The background field dependence enters in two places. First, the action  $S[\hat{\chi}; \bar{g}]$ includes gauge fixing and ghost terms and secondly  the cut-off $\mathcal{R}_k[\bar{g}]$ depends on covariant derivatives and a tensor structure   which are built from the background metric.
Similarly to the case of the scalar field, it follows from~\eqref{EAA_def} that
\beq
f = \langle \hat{f}_k \rangle_{f,k} \,,
\eeq
where the expectation value of any functional of the fields $\hat{\mathcal{O}}[\hat{\chi}]$ is defined by
\beq
\langle \hat{\mathcal{O}} \rangle_{f,k} :=  {\rm{e}}^{\Gamma_k[f,\bar{g}]} \int \left({\rm{d}} \hat{\chi}\right)\, {\rm{e}}^{-S[\hat{\chi}; \bar{g} ]} {\rm{e}}^{(\hat{f}_k[\hat{\chi}] - f)\cdot \frac{\delta }{\delta f}  \Gamma_k[f;\bar{g}] }
{\rm{e}}^{- \frac{1}{2} (\hat{f}_k[\hat{\chi}] - f)\cdot  \mathcal{R}_k[\bar{g}] \cdot  (\hat{f}_k[\hat{\chi}] - f) }  \hat{\mathcal{O}}[\hat{\chi}] \,.
\eeq
The generalised flow equation for $\Gamma_k[f;\bar{g}]$ is given by
\begin{equation} \label{dimensionfull_flow}
\left( \partial_t  +   \Psi_k[f;\bar{g}] \cdot  \frac{\delta }{\delta f}  \right) \Gamma_k[f;\bar{g}]
 = \frac{1}{2} \STr \,  \mathcal{G}_k[f;\bar{g}]  \left( \partial_t  + 2  \cdot \frac{\delta}{\delta f}  \Psi_k[f;\bar{g}] \right)\cdot \mathcal{R}_k[\bar{g}] \,,
\end{equation}
where $f = \langle \hat{f} \rangle$ are the mean fields and $\mathcal{G}_k[f,\bar{g}]$ denotes the propagator
\beq
\mathcal{G}_k[f;\bar{g}] :=   \left(\frac{\delta}{\delta f } \Gamma_k[f;\bar{g}]  \frac{\overleftarrow{\delta}}{  \delta f }+ \mathcal{R}_k[\bar{g}] \right)^{-1} \,,
\eeq
with $\overleftarrow{\delta}$ signifying that the derivative acts to the left.
The $\cdot$ implies a continuous matrix multiplication including sum over all field components and integration over spacetime. The $\STr$ denotes a supertrace in the same sense with a minus sign inserted for anti-commuting  fields.
 For gravity the RG kernel now has component for each field $\Psi_k = \{ \Psi_{\mu\nu}^g, \Psi^{c\mu}, \Psi^{\bar{c}}_\mu\}$, such that
$
 \Psi_k  = \langle  \partial_t \hat{f}_k \rangle_{f,k}
$\,.
By setting $\Psi_k =0$ we obtain the flow equation for gravity derived in~\cite{Reuter:1996cp}, however in this case we would have to also compute the flow of inessential couplings.
Using the background field method~\cite{Abbott:1981ke}, one is ultimately interested in identifying $\bar{g}_{\mu\nu} = g_{\mu\nu}$ and setting $c^\mu =0 = \bar{c}_\nu$. It is, therefore, convenient  to write the action as

\beq
\Gamma_k[g, c ,\bar{c};\bar{g}]= \bar{\Gamma}_k[g] + \hat{\Gamma}_k[g, c ,\bar{c};\bar{g}] \,,
\eeq
where
\beq
\bar{\Gamma}_k[g] \equiv  \Gamma_k[g, 0 ,0;g]  \implies   \hat{\Gamma}_k[g, 0 ,0;g]=0
\eeq
is a diffeomorphism invariant action and   $\hat{\Gamma}_k[g, c ,\bar{c};\bar{g}]$ contains terms which depend on the ghosts and the two metrics separately, including the ghost and gauge fixing terms.   
The diffeomorphism invariant action has the derivative expansion
\beq \label{barGamma}
\bar{\Gamma}_k[g] =   \int \d^dx  \sqrt{\det g}\left\{\frac{\rho_k}{8 \pi}    - \frac{1}{16 \pi G_k} R  + a_k R^2  + b_k R_{\mu\nu} R^{\mu\nu} + c_k  E  + O(\partial^6) \right\} \,.
\eeq
Here $G_k$ and $\rho_k$ are the running Newton's constant and vacuum energy, respectively, and $a_k$, $b_k$ and $c_k$ multiply the $O(\partial^4)$ terms with  $E = R_{\mu \nu \a \b}R^{\mu \nu \a \b}-4 R_{\mu \nu}R^{\mu \nu}+ R^2$. It will also be useful to define the cosmological constant as
\beq
\Lambda_k := \rho_k G_k \,,
\eeq
since it is this combination that appears in the canonically normalised propagator.
In four dimensions the integral  $ \int \d^4x  \sqrt{\det g} E$ is a topological invariant, so $c_k$ will not enter into any derivative of $  \bar{\Gamma}_k[g] $ and, as such, $c_k$ does not appear in any beta function~\cite{Falls:2020qhj,Knorr:2021slg}.

At a non-trivial fixed point required by asymptotic safety, the RG flow of dimensionless couplings in units of $k$ will become independent of $k$.
As such, it is convenient to define the dimensionless couplings
\beq
\tilde{G}(t) := k^{d-2} G_k \,, \,\,\,\,\,\,\,  \tilde{\rho}(t) := k^{-d} \rho_k\,   \,,\,\,\,\,\,\,  \tilde{\Lambda}(t):= \tilde{G}(t) \tilde{\rho}(t) \,,
\eeq
where we will omit to make the $t$-dependence of the dimensionless couplings explicit in the following.

Here we shall use the commonly used background field approximation where \linebreak 
$\hat{\Gamma}_k[g, c ,\bar{c};\bar{g}]$ is approximated by a BRST invariant action consisting of the background covariant gauge fixing and ghost terms. In particular, we shall take
\beq
\hat{\Gamma}_k[g, c ,\bar{c};\bar{g}] = \frac{1}{2 } \int \d^d x \sqrt{\det \bar{g}} \left( F^{\nu}  \bar{g}_{\mu\nu} F^{\mu}   + \bar{c}_\mu \mathcal{Q}^{\mu}\,_{\nu} c^{\nu} \right) \,,
\eeq
where, to simplify 
 calculations,  we adopt background covariant harmonic gauge
\bea
F^\mu
&=& \frac{\sqrt{2}}{\kappa_k}  \left(  \bar{g}^{\mu \lambda} \bar{g}^{ \nu \rho}  - \frac{1}{2} \bar{g}^{\nu\mu}   \bar{g}^{\rho \lambda}     \right)  \bar{\nabla}_\nu  g_{\lambda \rho}\, ,
\eea
and $\kappa_k$ denotes  the dimensionful coupling
\beq
\kappa_k \equiv \sqrt{32 \pi G_k}\,\,.
\eeq
The ghosts operator is then given by
\beq
\mathcal{Q}^{\mu}\,_{\nu} c^{\nu} \equiv \mathcal{L}_{c} F^{\mu} =   \frac{\sqrt{2}}{\kappa_k}  \left(  \bar{g}^{\mu \lambda} \bar{g}^{ \nu \rho}  - \frac{1}{2} \bar{g}^{\nu\mu}   \bar{g}^{\rho \lambda}     \right)  \bar{\nabla}_\nu ( g_{ \rho \sigma} \nabla_{\lambda} c^{\sigma}  + g_{ \lambda \sigma} \nabla_{\rho} c^{\sigma}) \,.
\eeq
In the background field approximation, we will choose $ \Psi^{c\mu} = 0 = \Psi^{\bar{c}}_\mu$, while we choose the RG kernel for the metric to be given by
\beq \label{Psi_expansion}
\Psi_{\mu\nu}^g[g] = \gamma_g g_{\mu\nu} + \gamma_R R g_{\mu\nu} +  \gamma_{Ricci}   R_{\mu\nu} + O(\partial^4) \,,
\eeq
where $\gamma_i$ with $i =\{ g, R, Ricci \}$ are the `gamma functions' which, along with the beta functions, will be determined as functions of the couplings that appear in $\bar{\Gamma}_k[g]$. Each gamma function allows us to impose a renormalisation condition which fixes the flow of an inessential coupling. Thus, retaining three gamma functions allows us to impose three renormalisation conditions which are constraints on the form of $\bar{\Gamma}_k[g]$ that we impose along the RG flow. We note that $\gamma_g $ is dimensionless while $\gamma_R$ and $\gamma_{Ricci}$ have mass dimension $-2$, thus we define dimensionless gamma functions $\tilde{\gamma}_{R} := k^2 \gamma_{R}$ and  $\tilde{\gamma}_{Ricci} := k^2 \gamma_{Ricci}$.
As with the derivative expansion for a scalar field, if we work at order $\partial^s$ in the derivative expansion, we include all terms of order $\partial^{s-2}$ in the RG kernel~\eqref{Psi_expansion}.

In our approximation the flow for the diffeomorphism invariant action $\bar{\Gamma}_k[g]$ is given~by
\begin{equation} \label{dimensionfull_flow_BF}
\left( \partial_t  +   \Psi_k^g \cdot  \frac{\delta }{\delta g}  \right) \bar{\Gamma}_k
= \frac{1}{2} \Tr \,  \mathcal{G}_k^{gg}  \left( \partial_t  + 2  \cdot \frac{\delta}{\delta g}  \Psi_k^g \right)\cdot \mathcal{R}_{k}^{gg}  - \Tr \,  \mathcal{G}_k^{\bar{c}c } \cdot \partial_t \mathcal{R}^{\bar{c} c}_k \,,
\end{equation}
where
\begin{align}
& \mathcal{G}_k^{\bar{c}c } := \frac{1}{ K_{\bar{c} c} \cdot \Delta_{\rm gh}  + \mathcal{R}_k^{\bar{c}c }} \,,
\\
& \mathcal{G}_k^{gg} :=  \frac{1}{ \frac{\delta^2 \bar{\Gamma}_k}{\delta g \delta g} +   K_{g g} \cdot \Delta_{\rm gf}  + \mathcal{R}^{gg}_k} \,,
\end{align}
$\Delta_{\rm gh}$ and $\Delta_{\rm gf}$ denote the differential operators
\bea
\Delta_{\rm gh}\,^\mu\,_\nu &=& - \delta^\mu_\nu \nabla^2  - R^{\mu}\,_\nu \,,
\,\,\,\,\,\,\,\,\,\,\,\,\,
(\Delta_{\rm gf})_{\mu\nu}\,^{\rho \lambda}   = \nabla_{\mu} \nabla_{\nu}   g^{\rho \lambda}  - 2 \delta^{(\rho}_{(\nu} \nabla_{\mu)} \nabla^{\lambda)}\,,
\eea
and
\bea
K^{\mu\nu, \a\b}_{gg} &:=& \frac{1}{ 2 \kappa^2_k} \,  \sqrt{\det g} \left( g^{\mu \a}g^{\nu \b}+g^{\mu \b}g^{\nu \a}-g^{\mu \nu}g^{\a \b} \right)\,,  \,\,\,\,\,\,\,\,\,
K^{\mu\nu}_{\bar{c} c} :=\frac{\sqrt{2} }{\kappa_k} \sqrt{\det g} \,g_{\mu \nu}\,.
\eea
We then choose the regulators to be of the form
\begin{align}
\mathcal{R}_k^{gg}[g] = K_{gg} R_k(\Delta) \,,
\hspace{3cm}
\mathcal{R}_k^{\bar{c} c}[g] = K_{c\bar{c}}  R_k(\Delta) \,,
\end{align}
where $\Delta = - g^{\mu\nu}\nabla_\mu \nabla_\nu$ is the Laplacian.

The redundant operators for $\bar{\Gamma}_k[g]$ are given by
\begin{equation} \label{Redundant_BF}
\mathcal{T}[\Phi^g] :=  \Phi^g \cdot  \frac{\delta }{\delta g}  \bar{\Gamma}_k
-\Tr \,  \mathcal{G}_k^{gg}     \cdot \frac{\delta}{\delta g}  \Phi^g \cdot \mathcal{R}_{k}^{gg} \,,
\end{equation}
where $\Phi^g $ are symmetric covariant tensors composed of the metric, curvature tensors and their covariant derivatives, e.g., $\Phi^g_{\mu\nu} = g_{\mu\nu}, R g_{\mu\nu}, R_{\mu\nu}$.

The minimal essential scheme for quantum gravity, which we will further elaborate on in Sections~\ref{sec:vaceniness} and ~\ref{sec:minessschemequagra}, closely follows the perturbative scheme put forward in~\cite{Anselmi:2002ge,Anselmi:2013wha}. The scheme puts to zero any term that vanishes when the vacuum Einstein equations
\beq \label{VacuumEinstein}
R_{\mu\nu} = 0
\eeq
apply apart from the Einstein--Hilbert term itself. The reasoning is that the fixed point where $\tilde{G}=0$ and $\tilde{\Lambda} =0$ is the analog of the GFP for a scalar field theory.
 This means that can we set to zero both $a_k=0$ and $b_k=0$, while leaving $c_k$ non-zero since this term is topological in $d=4$.
As with the GFP~\eqref{Scalar_GFP} in the scalar field theory the fact that any operator that vanishes on the equations of motion~\eqref{VacuumEinstein} can be removed by a field redefinition is a property of the fixed point where $\tilde{G}=0$ and $\tilde{\Lambda} =0$.
 A higher derivative Gaussian fixed point, more analogous to the fourth order fixed point~\eqref{Scalar_GFP_4}, is achieved by instead writing 
 \beq
 a_k = - \frac{ 1+ \omega}{ 3 \lambda} \,,   \hspace{2cm}  b_k = \frac{1}{\lambda} \,,  \hspace{2cm} c_k = \frac{ 1- 2\theta}{ 2 \lambda} \,,
 \eeq
 and sending $\lambda \to 0$. At this fixed point the degrees of freedom are those of Stelle's higher derivative gravity rather than Einstein gravity.
 Furthermore, since the equations of motion for higher derivative gravity do not imply~\eqref{VacuumEinstein} the couplings $a_k$ and $b_k$ (or equivalently $\lambda$ and $\omega$) are essential at the higher derivative Gaussian fixed point.

Here we concentrate on Einstein gravity where $a_k$ and $b_k$ are inessential in the vicinity of the GFP where $\tilde{G}=0$ and $\tilde{\Lambda} =0$.  Thus, after setting $a_k=0$ and $b_k=0$ and neglecting all terms with more than four derivatives in $\bar{\Gamma}_k$, while retaining $\gamma_g$, $\gamma_R$, and $\gamma_{Ricci}$ we expand the Equation~\eqref{dimensionfull_flow} to order $\partial^4$ to obtain five flow equations from the independent tensor structures present in~\eqref{barGamma} using off-diagonal heat kernel techniques~\cite{Groh:2011dw}. The evaluation of the traces and the resulting flow equations are presented in appendix~\ref{app:gravitytrace}. The equations are presented for arbitrary cut-off function $R_k(\Delta)$ and in arbitrary dimension $d$ neglecting terms proportional $c_k$ in the traces: this is justified in $d=4$ since in this case the corresponding invariant is topological. For the remainder of the paper we will take $d=4$.
For explicit calculations, we will use the Litim cut-off function
\beq
R_k(\Delta) = (k^2 - \Delta) \Theta(k^2 - \Delta) \,,
\eeq
where $\Theta(x)$ is the Heaviside theta function.
%
\section{The Vacuum Energy Is Inessential}
\label{sec:vaceniness}
Having set $a_k$ and $b_k$ to zero, we can solve the equations for $\gamma_R$ and $\gamma_{Ricci}$, given in~\eqref{eq:gammarsca} and~\eqref{eq:gammaricci}.
What is less clear is which renormalisation condition we should apply to freeze the inessential coupling associated to $\gamma_g$, that must be some combination of $G_k$ and $\rho_k$. As has been pointed out in~\cite{Percacci:2004sb}, to find a non-trivial fixed point in gravity must actually require $\tilde{G}$ to have a fixed point. Indeed, by rescaling the metric, or, in other words, choosing a system of units, one cannot set $G_k=1$ and $k=1$ simultaneously.  This is still the case even with $\gamma_g$ present since one does not find a non-trivial fixed point for $\tilde{\Lambda}$ if we try to fix the condition that $G_k = G_0$. The reason is that the beta function for $\tilde{\Lambda}$ still depends on $k^2 G_0$ which diverges as $k \to \infty$. However, one still has the freedom to apply one RG condition afforded by the presence of $\gamma_g$. What is evident is that the dimensionless inessential coupling will still need a fixed point value. Thus, we should instead fix a dimensionless coupling to one value along the RG flow. However, one finds that doing so can prevent the GFP from being present itself. For example if we set $\tilde{G}=1$ or $\tilde{\Lambda}=1$, the GFP,  which in dimensionless variables is at  $\tilde{G}=0$ and $\tilde{\Lambda}=0$, cannot be attained.
 This is a consequence of the fact that with a specific renormalisation condition we cannot explore all universality classes contained in the theory space. In particular, since we will consider trajectories inside the subspace of theory space which contains the GFP, we will take into account the values of $\tilde G$ and $\tilde \Lambda$ at the GFP. Therefore,
to determine which dimensionless coupling we should fix, we analyse the GFP to understand which particular combination $\tilde{G}$ and $\tilde{\Lambda}$ is inessential.
However, we should understand this limit as the approach to a free theory on flat spacetime where $\bar{\Gamma}_k[g]$ reduces to the linearised Einstein--Hilbert action.  To see this limit properly we have to decompose the metric as\footnote{We stress that this decomposition  is independent of the split $g_{\mu\nu} = \bar{g}_{\mu\nu} + h_{\mu\nu}$, which is purely technical. }
\beq \label{ghat_to_phi}
\hat{g}_{\mu\nu} =   \mathfrak{g}_{\mu\nu} + \kappa_k \hat{\phi}_{\mu\nu}\,,
\eeq
where $\mathfrak{g}_{\mu\nu}$ is a flat metric.
We call $\hat{\phi}_{\mu\nu}$ the graviton field since it is a fluctuation around a flat metric $\mathfrak{g}_{\mu\nu}$, allowing one to define asymptotic states as free gravitons. In the parameterisation~\eqref{ghat_to_phi}, it becomes clear that  $ \kappa_k$  is the coupling constant that measures the strength of self interactions of the graviton. The GFP corresponds to the theory where $\kappa_k = 0$. As we shall show later $\gamma_R$ and $\gamma_{Ricci}$ are both proportional to $\kappa_k^2$.
Taking a derivative of~\eqref{ghat_to_phi} with respect to $t$, we obtain
\beq
\partial_t\hat{g}_{\mu\nu}  =       \frac{1}{2} \eta_N   \kappa_k \hat{\phi}_{\mu\nu}  +     \kappa_k \partial_t \hat{\phi}_{\mu\nu}  + O(\kappa^2_k) \,,
\eeq
where  $\eta_N :=  \partial_t \log G_k $. The factor of $\kappa_k$ ensures that the field $\hat{\phi}_{\mu\nu}$ is canonically normalised.
 The expectation value of $\partial_t \hat{\phi}_{\mu\nu}$ is, therefore, given by
\beq
\Psi^\phi_{\mu\nu} \equiv \langle  \partial_t \hat{\phi}_{\mu\nu}(k)   \rangle  =  \frac{1}{ \kappa_k } \Psi_{\mu\nu}^g   - \frac{1}{2} \eta_N \phi_{\mu\nu}  + O(\kappa_k) \,,
\eeq
where $\phi_{\mu\nu} = \langle \hat{\phi}_{\mu\nu} \rangle$,  and inserting the expression for $\Psi^g$ we obtain
\beq \label{Psi^phi}
\Psi^\phi_{\mu\nu}  = \gamma_{\rm shift}  \mathfrak{g}_{\mu\nu}   - \frac{1}{2} \eta_\phi \phi_{\mu\nu} + O(\kappa_k)
\,,
\eeq
where
\beq
\eta_\phi :=\eta_N - 2 \gamma_g  \,,   \,\,\,\,\,\,\,\,  \gamma_{\rm shift} :=    \frac{\gamma_g}{ \kappa_k }
\eeq
are the anomalous dimension of the graviton field and the gamma function related to a shift of the graviton field by a constant. Imposing that  $\gamma_{\rm shift} $ is finite when $\kappa_k = 0$, we deduce that $ \gamma_g = 0$ at the GFP.
Defining
\beq
K_{\phi\phi} := \kappa_k^2 K_{gg},\,\,\,\,\,\,\,\,\, \mathcal{R}^{\phi\phi}_k :=  \kappa_k^2  \mathcal{R}^{g g}_k \,,
\eeq
the flow Equation~\eqref{dimensionfull_flow} can be rewritten as
\begin{equation} \label{dimensionfull_flow_phi}
\left( \partial_t|_{\phi} +   \Psi_k^\phi \cdot  \frac{\delta }{\delta \phi}  \right) \bar{\Gamma}_k
= \frac{1}{2} \Tr \,  \mathcal{G}_k^{\phi\phi}  \left( \partial_t  + 2  \cdot \frac{\delta}{\delta \phi}  \Psi_k^\phi \right)\cdot \mathcal{R}_{k}^{\phi\phi}  - \Tr \,  \mathcal{G}_k^{\bar{c}c } \cdot \partial_t \mathcal{R}^{\bar{c} c}_k \,,
\end{equation}
where the canonically normalised regularised propagator is
\beq
\mathcal{G}_k^{\phi\phi} =  \frac{1}{ \frac{\delta^2 \bar{\Gamma}_k}{\delta \phi \delta \phi} +   K_{\phi \phi} \cdot \Delta_{\rm gf}  + \mathcal{R}^{\phi\phi}_k} \,.
\eeq
Inserting $g_{\mu\nu} =    \mathfrak{g}_{\mu\nu} + \kappa_k \phi_{\mu\nu}$ into $\bar{\Gamma}_k[g]$ and then expanding in $\kappa_k$,  we find that at the GFP the EAA has the form
\beq \label{GFP}
\bar{\Gamma}^{\rm GFP}_k := \frac{1}{2} \phi\cdot   K_{\phi \phi}[\mathfrak{g}] \cdot (\Delta[\mathfrak{g}] - \Delta_{\rm gf} [\mathfrak{g}] ) \cdot   \phi  +  k^4  \frac{1}{8 \pi} \int \d^4 x \sqrt{\det \mathfrak{g}}  \tilde{\rho}_{\rm GFP} \,,
\eeq
where we anticipate that for $\kappa_k =0$ the vacuum energy is $\rho_k = k^4 \tilde{\rho}_{\rm GFP}$ and $ \tilde{\rho}_{\rm GFP}$ denotes the dimensionless fixed point value for the vacuum energy, which we will determine shorty. Inserting~\eqref{GFP} into the LHS of~\eqref{dimensionfull_flow_phi} we obtain
\beq \label{LHS_GFP_eq}
\left( \partial_t|_{\phi} +   \Psi_k^\phi \cdot  \frac{\delta }{\delta \phi}  \right) \bar{\Gamma}_{GFP} = \frac{1}{2 \pi }  k^4  \int \d^4 x \sqrt{\det \mathfrak{g}}  \tilde{\rho}_{\rm GFP}      - \frac{1}{2} \eta_\phi   \phi\cdot   K[\mathfrak{g}] \cdot (\Delta[\mathfrak{g}] - \Delta_{\rm gf} [\mathfrak{g}] ) \cdot \phi \,,
\eeq
while on the RHS we have, using that $\gamma_g =0$,
\beq
\frac{1}{2} \Tr \,  \mathcal{G}_k^{\phi\phi}  \left( \partial_t  + 2  \cdot \frac{\delta}{\delta \phi}  \Psi_k^\phi \right)\cdot \mathcal{R}_{k}^{\phi\phi}  - \Tr \,  \mathcal{G}_k^{\bar{c}c } \cdot \partial_t \mathcal{R}^{\bar{c} c}_k  = k^4 \int_0^\infty \d z\, z \frac{ -3 \eta_\phi  R_k(z)+\partial_t R_k(z)}{16 \pi ^2
   (R_k(z)+z)} \,,
\eeq
which is independent of $\phi_{\mu\nu}$ and, as such, we find that $\eta_\phi=0$ at the GFP which together with $\gamma_g=0$ implies $\eta_N= 0$.
We then see that the GFP value of the dimensionless vacuum energy is
\beq
\tilde{\rho}_{\rm GFP} = \frac{1}{8 \pi }  \int_0^\infty \d z  \,z \,\frac{\partial_t R_k(z)}{z + R_k(z)}   \,.
\eeq
Using the Litim cut-off we obtain the value
\beq
\tilde{\rho}_{\rm GFP} =  \frac{1}{8 \pi } \,.
\eeq
We conclude that the GFP is characterised uniquely by $\tilde{G}=0$, $\eta_N = 0$, $\gamma_g=0$ and a scheme dependent value for the dimensionless vacuum energy $\tilde{\rho} = \tilde{\rho}_{\rm GFP}$. The fact that $\eta_N = 0$ means that we arrive at the GFP in dimensionless variables when $G_k \to G_0$ is a constant, such that $\tilde{G}$ vanishes as $ \tilde{G} \sim k^2 G_0$ in the limit $k\to 0$.
Thus the GFP is an IR fixed point for~$\tilde{G}$.

Two remarks are in order concerning the vacuum energy.
First, let us note that we could also choose a more general cut-off scheme allowing for different cut-off functions for the ghosts and gravitons   in such a manner that $\tilde{\rho}_{\rm GFP}$ would vanish~\cite{Falls:2017cze}.
At the exact level no physics should depend on the choice of cut-off so the value of $\tilde{\rho}_{\rm GFP}$ should be of no significance. Secondly, we note that it may seem we could satisfy the flow with $\rho_k =  k^4  \tilde{\rho}_{\rm GFP} + \rho_0$ allowing for a non zero cosmological constant, since $\rho_0$ is a constant of integration that will not appear in~\eqref{LHS_GFP_eq}. However, only with $\rho_0= 0$ do we have a  fixed~point. 

Now, away from the GFP, $\gamma_g$ needs not be equal to zero,
so we can now write the linearised ansatz for $\gamma_g$ around the GFP as
\beq
\gamma_g = w_1 \left( \tilde{\rho} -\frac{1}{8 \pi} \right)      +  w_2 \tilde{G} + \dots \,,
\eeq
where $w_1$ and $w_2$ are free parameters which we are free to choose and the dots are non-linear  terms in the expansion around the GFP. Expanding the beta functions for $\tilde{G}$ and $\tilde{\rho}$ we obtain
\bea
\partial_t \tilde{G} &=& 2 \tilde{G} + \dots \,,\\
\partial_t \tilde{\rho} &=& \left(\frac{w_1}{3 \pi }-4\right) \left(\tilde{\rho} -\frac{1}{8 \pi }\right) + \left( \frac{w_2}{3 \pi }+\frac{38}{24 \pi ^2} \right) \tilde{G} +   \dots
\label{linear_beta_rho}
\eea
and, thus, we see that the linearised flow of $\tilde{G}$ around the GFP is scheme independent, while the linearised flow of $\tilde{\rho}$ is scheme dependent.
Since scheme dependence is the hallmark of an inessential coupling, we can conclude that Newton's coupling $\tilde{G}$ is an essential coupling in the vicinity of the GFP, while $\tilde{\rho}$ is inessential.
We are free to specify the flow for $\tilde{\rho}$ instead of computing it and we can freely choose the corresponding scaling dimension rather than assuming it should have dimension $4$. In fact, we can even make the vacuum energy, which canonically is the most relevant coupling, an irrelevant coupling simply by choosing  $w_1 > 12 \pi$. Let us stress that these are exact statements since we are at the GFP and terms at order $\partial^6$ arise at two loops.

A remarkable consequence of the vacuum energy being inessential is that we may simply choose that $\rho_{k=0} = 0$ and, thus, the vanishing of the vacuum energy is achieved by a renormalisation condition. Thus, at least in pure gravity, there is no fine tuning problem related to the cosmological constant once we apply this condition. However, this condition dictates the vanishing of the cosmological constant and by imposing it we are restricting which theories we can have access to. This suggests that there is a universality class of quantum gravity where the cosmological constant is zero. This universality class possesses the IR GFP where $G_0$ is a constant and $\rho_0 =0$. Although there may be other universality classes where the cosmological constant is non-zero, here we will explore this one to see if there is also a non-trivial fixed point that can be used to define the interacting QFT.

Before ending this section, let us stress two points regarding the interpretation of couplings in gravity and the possible (imperfect) analogies we can make with couplings in,  e.g., $\phi^4$-theory. First, despite appearances, $G_k$ is not the inverse wave-function renormalisation, but a  coupling, more analogous to the interaction coupling $\lambda$ in $\phi^4$-theory.
In particular, while the wave-function renormalisation is an inessential coupling in  $\phi^4$-theory, $G_k$ is an essential coupling like $\lambda$.
Secondly, again despite appearances, $\Lambda_k =\rho_k G_k$ is not a mass squared. A more clear interpretation of the vacuum energy comes if we choose to parameterise the metric, such that $\rho_k \sqrt{\det g}$ is linear in the field $\sigma$ which parameterises conformal fluctuations~\cite{Falls:2015qga}.
This can be achieved by setting  $g_{\mu\nu} = \left( 1+ \frac{\sigma}{d}    \right)^{\frac{2}{d}} \mathfrak{g}_{\mu\lambda} (e^{h})^\lambda\,_\nu$, where $h_{\mu}^{\mu} = 0$, such that
$\rho_k \sqrt{\det g} =  \rho_k   \sqrt{\det \mathfrak{g}}  \left(1+ \frac{\sigma}{d} \right)$ is linear in $\sigma$. The fact that $\rho_k$ couples also to the purely vacuum term is what is crucial for $\rho_k$  to be inessential. Thus $\rho_k$, rather than being analogous to a mass in a scalar theory which is essential, can better be interpreted as a constant source which couples linearly to the field in the broken phase of diffeomorphisms.
\section{Minimal Essential Scheme for Quantum Einstein Gravity}
\label{sec:minessschemequagra}
Since the vacuum energy is inessential coupling at the GFP, we can fix it by a renormalisation condition. In particular, we can pick a condition which ensures that we are in the universality class which possesses the GFP and removes the vacuum energy from the set of couplings we must compute the flow of. We will adopt the simplest RG condition of this type which sets
\beq \label{RG_condition}
\tilde{\rho}(t)\equiv k^{-4} \rho_k = \tilde{\rho}_{\rm GFP}
\eeq
for all scales $k = k_0 e^{t}$.  The RG condition~\eqref{RG_condition} identifies the vacuum energy with cut-off scale $\rho_k = k^4 \tilde{\rho}_{\rm GFP}$. Having applied~\eqref{RG_condition}, then the dimensionless product $\tau_k := \rho_k G^2_k$ is given by
\beq
\tau_k = \tilde{\rho}_{\rm GFP} \tilde{G}(t)^2
\eeq
and, therefore, the flow of $\tilde{G}(t)$ completely determines the flow of $\tau_k$.
In classical general relativity, in the absence of matter, $\tau_k$ is the only meaningful coupling since one can rescale the metric.
This can be seen explicitly from the flow equation by observing that, when the RHS is neglected, the beta function for $\tau_k$ is independent of $\gamma_g$. More generally, when $k=0$ it is evident that only dimensionless ratios couplings can be essential since a rescaling of the metric will change the values of dimensionful couplings. As such, $\tau_0$ is the physical cosmological constant in Planck units which can be considered as an observable,  which vanishes in the universality class we are considering.

Let us stress, however, that although $\rho_k$ will vanish at $k=0$, its presence in the action is still needed to consistently solve the flow equation for non-zero $k$. If we would simply neglect the flow of $\rho_k$ entirely, then $G_k$ would appear to be inessential since we could instead use $\gamma_g$ to dictate the flow of $G_k$ instead.

In addition to~\eqref{RG_condition}, we specify an infinite set of renormalisation conditions which exclude all terms that are dependent on the Ricci curvature $R_{\mu\nu}$ from the ansatz from $\bar{\Gamma}_k$ apart from the Einstein--Hilbert action and the topological Gauss--Bonnet term. This defines the minimal essential scheme for quantum gravity. At second order in curvature, the most general  diffeomorphism invariant action can be written as
\vspace{-3pt}
 \beq
  \bar{\Gamma}_k[g] =   \int \d^4x  \sqrt{\det g}\left\{\frac{\rho_k}{8 \pi}    - \frac{1}{16 \pi G_k} R  + R W_{R,k}(\Delta) R + R_{\mu\nu} W_{Ricci,k}(\Delta)R^{\mu\nu} + c_k  E  + O(R_{\mu\nu\rho\lambda}^3) \right\}
 \eeq
and, hence, in the minimal essential scheme we set $ W_{R,k}(\Delta) = 0 =  W_{Ricci,k}(\Delta)$. 
Since, furthermore, all the higher terms depend only on the Weyl curvature $C_{\mu\nu\rho\lambda}$, the propagator evaluated on any conformally flat spacetime, i.e., those where $C_{\mu\nu\rho\lambda}=0$,  is just that of classical general relativity~\cite{Anselmi:2013wha}. 
 Consequently, the regularised propagator  evaluated on a conformally flat background takes the form
\beq \label{minimal_propagator}
\mathcal{G}_{k}^{gg} =  K_{g g}^{-1} \cdot \frac{1}{\Delta  -2 k^4 G_k \tilde{\rho}_{\rm GFP} + R_k(\Delta)}+ O(R_{\mu\nu}) \,.
\eeq
This ensures that the theory at $k=0$ describes massless gravitons only.

Here, we shall only consider pure gravity. However, in~\cite{Anselmi:2002ge}, general arguments for spin $0$, $1/2$, $1$, $3/2$, and $2$ fields suggest that terms which would modify the propagator by introducing new poles are redundant in the vicinity of the GFP. For example if we consider a scalar tensor theory
\beq
\bar{\Gamma}_k[g,\phi] = - \frac{1}{16 \pi G_k} \int \d^4x \sqrt {\det g} \left[  - \frac{1}{16 \pi G_k} R  + \frac{1}{2} (\nabla_\mu \phi  \nabla^\mu \phi) + V_k(\phi) \right]\,,
\eeq
then we can still use both the equations of motion for the metric and the scalar to remove inessential couplings.
Out of all terms with up to four derivatives, the only additional terms which do not vanish when the equations of motion apply are 
\beq
\int \d^4x  \sqrt {\det g}  W_k(\phi)  (\nabla_\mu \phi  \nabla^\mu \phi)^2   +  \int \d^4x  \sqrt {\det g}  C_k(\phi) E \,,
\eeq
neither of which enter the propagator evaluated on a conformally flat spacetime and for constant values of $\phi$.

 The fact that we can remove the terms which lead to extra poles in the propagator along the RG flow indicates that the poles encountered in other schemes are spurious~\cite{DeAlwis:2018ybu}.
Let us stress however that nothing is wrong with using a scheme where the form factors do not vanish, such that the propagator at $k=0$ with $\rho_0 =0$ has the form
\beq
\mathcal{G}_{0} =  K_{g g}^{-1} \cdot \frac{1}{Z(\Delta) \Delta} + O(R_{\mu\nu\rho\lambda}) \,,
\eeq
where $ Z(\Delta)$ is a wave function renormalisation factor related to the form factors $W_{R,k}(\Delta)$ and  $W_{Ricci,k}(\Delta)$,
which have been computed in various approximations in~\cite{Denz:2016qks,Christiansen:2012rx,Christiansen:2014raa,Christiansen:2015rva,Knorr:2019atm,Knorr:2021niv} and the physical implications for scattering amplitudes have been discussed in~\cite{Draper:2020bop,Draper:2020knh,Platania:2020knd,Bonanno:2021squ}. (
In~principle there can be another independent wave function renormalisation related to the scalar degree of freedom that is introduced in theories such as $f(R)$ gravity. For simplicity, we discuss the case where there is only one wave function renormalisation which implies a linear relation between $W_{R,k}(\Delta)$ and  $W_{Ricci,k}(\Delta)$). There are two cases, either $ Z$ introduces new poles into the propagator, or it does not. In the latter case, we can remove $Z$ by a reparameterisation since it must be an entire function, and thus it is just a momentum-dependent wave function renormalisation. 
 In this case, we will find the same physics as in the  minimal essential scheme, namely, although the field redefinition would modify the vertices of the theory, the propagator would return to the minimal form~\eqref{minimal_propagator}. The case where $Z$ is not an entire function corresponds to a universality class not accessible to the minimal essential scheme for pure gravity. In particular, it would include particles other than the massless graviton. 
Thus, on one hand, the minimal essential scheme for quantum gravity, like its counterpart of scalar field theories~\cite{Baldazzi:2021ydj}, does put a restriction on what physics we can access by following the corresponding RG flow. On the other hand, this is a feature of the scheme, and not a bug, since the restricted theory space has a physical meaning, describing the interactions of gravitons which are fluctuation around a flat spacetime. 
Moreover, there is no reason why these fluctuations can not be strongly interacting, in particular $\tilde{G}$ can become of order unity.

It may of course be that this universality class, which only includes a massless graviton, does not contain a suitable UV fixed point and that one would need more degrees of freedom to describe a consistent theory of quantum gravity. For example, it could be the case that one would need the extra degrees of freedom which are present in higher derivative gravity and are needed to make the theory perturbatively renormalisable, or that one would need to add an $\sqrt{\det g} R^2$ which includes an extra scalar degree of freedom in addition to the graviton.
Here we will test the hypothesises that these extra degrees of freedom are not necessary for non-perturbative renormalisability.

\section{The Reuter Fixed Point in the Derivative Expansion}
\label{sec:reuterfixpoint}
To test the aforementioned hypothesis, the minimal essential scheme can be carried out at each order in the derivative expansion. Here we will study the RG flow at order $\partial^2$, where the action is the Einstein--Hilbert action with~\eqref{RG_condition}, and at order $\partial^4$ where the action takes the form
 \beq \label{Essential_Action}
  \bar{\Gamma}_k[g] =   \int \d^4x  \sqrt{\det g}\left\{  k^4 \frac{\tilde{\rho}_{\rm GFP}}{8 \pi}    - \frac{1}{16 \pi G_k} R  + c_k  E  + O(\partial^6) \right\} \,,
 \eeq
with the only order $\partial^4$ term being the topological one.
At order $\partial^2$ we set $\gamma_R$ and $\gamma_{Ricci}$ to zero, along with all higher-order terms in $\Psi_k$, and expand the flow equation for~\eqref{Essential_Action} to order
$\partial^2$ solving for
\beq
\gamma_g = \gamma_g( \tilde{G}) \,, \,\,\,\,\,\,\,\,\,    \partial_t \tilde{G}= \beta_{\tilde{G}}( \tilde{G}) \,,
\eeq
which are functions of $\tilde{G}$ alone.  At order  $\partial^4$  we include all order $\partial^4$ tensor structures in the flow equations but solve for $\gamma_R$ and $\gamma_{Ricci}$ instead of the running of the higher derivative couplings $a_k$ and $b_k$, which are set to zero.
Thus, at order $\partial^4$ the minimal essential flow is characterised by five dimensionless functions of $\tilde{G}$, namely
 \bea
\gamma_g &=& \gamma_g( \tilde{G}) \,, \,\,\,\,\,    \partial_t \tilde{G}= \beta_{\tilde{G}}( \tilde{G}) \,,\\ [5pt]
\tilde{\gamma}_R &=& \tilde{\gamma}_R( \tilde{G}) \,, \,\,\,\,\,   \tilde{\gamma}_{Ricci} = \tilde{\gamma}_{Ricci}( \tilde{G})  \,, \,\,\,\,\,  \partial_t c_k= \beta_c(\tilde{G}) \,.
\eea
Let us stress that calculation is vastly simpler than the calculation where higher derivative couplings $a_k$ and $b_k$ do not vanish~\cite{{Falls:2020qhj,Knorr:2021slg}} and that the final form of the beta and gamma functions only depend on one coupling rather than four in the standard scheme.

As a first check we can analysis the behaviour around the GFP at $G_k=0$ to see how the universal one-loop divergencies are accounted for. In particular, the one-loop divergencies encountered in dimensional regularisation in our chosen gauge with $\Lambda =0$  are given by three terms
\beq \label{One-Loop-dim-Reg}
\Gamma_{\rm div} =  \frac{1}{d-4}\frac{1}{\left(4\pi\right)^2} \int \d^dx \sqrt{\det g}
\left[
\frac{1}{60}  R^2  + \frac{7}{10} R_{\mu\nu}  R^{\mu\nu} +  \frac{53}{45}  E
\right] \,.
\eeq
Upon replacing $ \frac{1}{d-4} \to \log(k/k_0)$ and taking a derivative with respect to $k$ the same three terms will appear in the flow equation on the RHS of Equations~\eqref{eq:gammarsca}--\eqref{eq:flowc}, respectively.
 However, the terms that would renormalise $a_k$ and $b_k$ are, instead, absorbed into $\gamma_{R}$ and  $\gamma_{Ricci}$, while $c_k$ will still be renormalised. Indeed, expanding in $G_k$ we find that
\beq
\gamma_{R} = - \frac{11}{30 \pi} G_k + O(G^2_k)\,,   \,\,\,\,\,   \gamma_{Ricci} = \frac{7}{10 \pi} G_k + O(G^2_k)\,, \,\,\,\,\, \beta_c = \frac{1}{(4 \pi)^2} \frac{53}{45} + O(G_k) \,,
\eeq
which precisely account for the divergences~\eqref{One-Loop-dim-Reg}.

\begin{figure}[h]
\includegraphics[scale=.8]{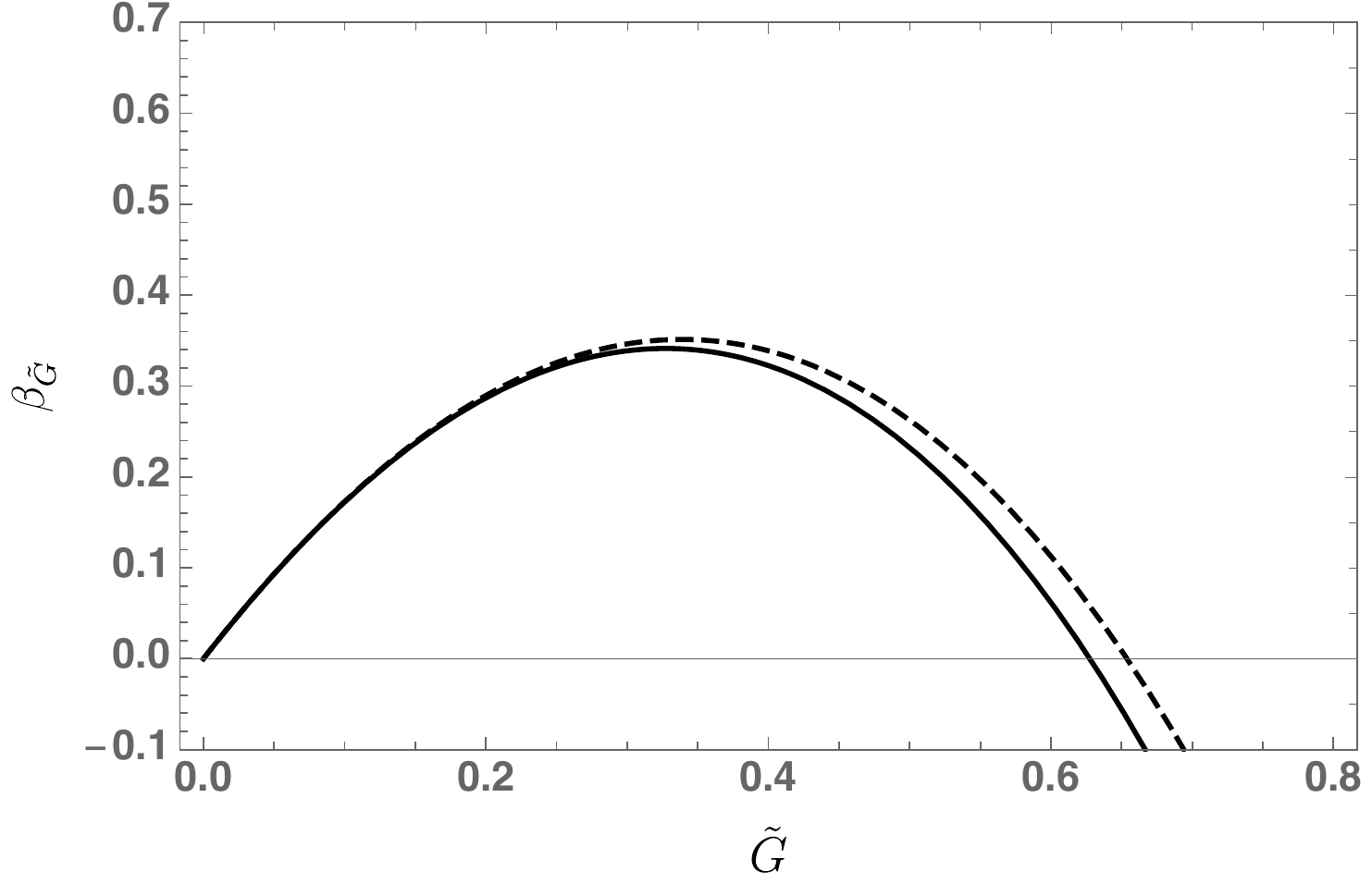}
\caption{The beta function for Newton's constant in the Einstein-Hilbert approximation (dashed line) and the order $\partial^4$ approximation (solid line). }
\label{fig:Beta_G}
\end{figure}

The non-perturbative beta functions $\beta_{\tilde{G}}( \tilde{G})$ at orders $\partial^2$ and $\partial^4$ are plotted in   Figure~\ref{fig:Beta_G} and are seen to closely agree for $\tilde{G}$ in the plotted region. At both orders there exists a UV fixed point where \bea \tilde{G} &=&\tilde{G}_\star = 0.6538\, \,\,\,\,\, {\rm at } \,\, O(\partial^2)\,,  \vspace{1cm} \\    \tilde{G}&=&\tilde{G}_\star =0.6275  \,\,\,\,\,\, {\rm at } \,\, O(\partial^4) \,,\eea which we can identify as the Reuter fixed point~\cite{Reuter:1996cp,Souma:1999at}.
The Reuter fixed point splits the phase diagram of quantum gravity into a weakly coupled and strongly coupled regions for $0<\tilde{G} < \tilde{G}_{\star}$ and $\tilde{G} > \tilde{G}_{\star}$, respectively.
The critical exponent at the Reuter fixed point   
\beq
\uptheta = - \frac{\partial \beta_{\tilde{G}}}{\partial  \tilde{G}}(\tilde{G}_\star) 
\eeq
 is given by 
 \bea
\uptheta &=& 2.3129 \,\,\,\,\,\, {\rm at }  \,\, O(\partial^2)\,,  \vspace{1cm} \\  
\uptheta &=& 2.3709  \,\,\,\,\,\, {\rm at }  \,\, O(\partial^4)\,,
\eea
which can be compared to the canonical scaling dimension of $\uptheta_{\rm can} = 2$ which is obtained at one-loop, and, therefore, receives a small correction. This suggests that the Reuter fixed point is weakly non-perturbative~\cite{Falls:2013bv,Eichhorn:2018ydy}.

\begin{figure}[h]
\includegraphics[scale=.8]{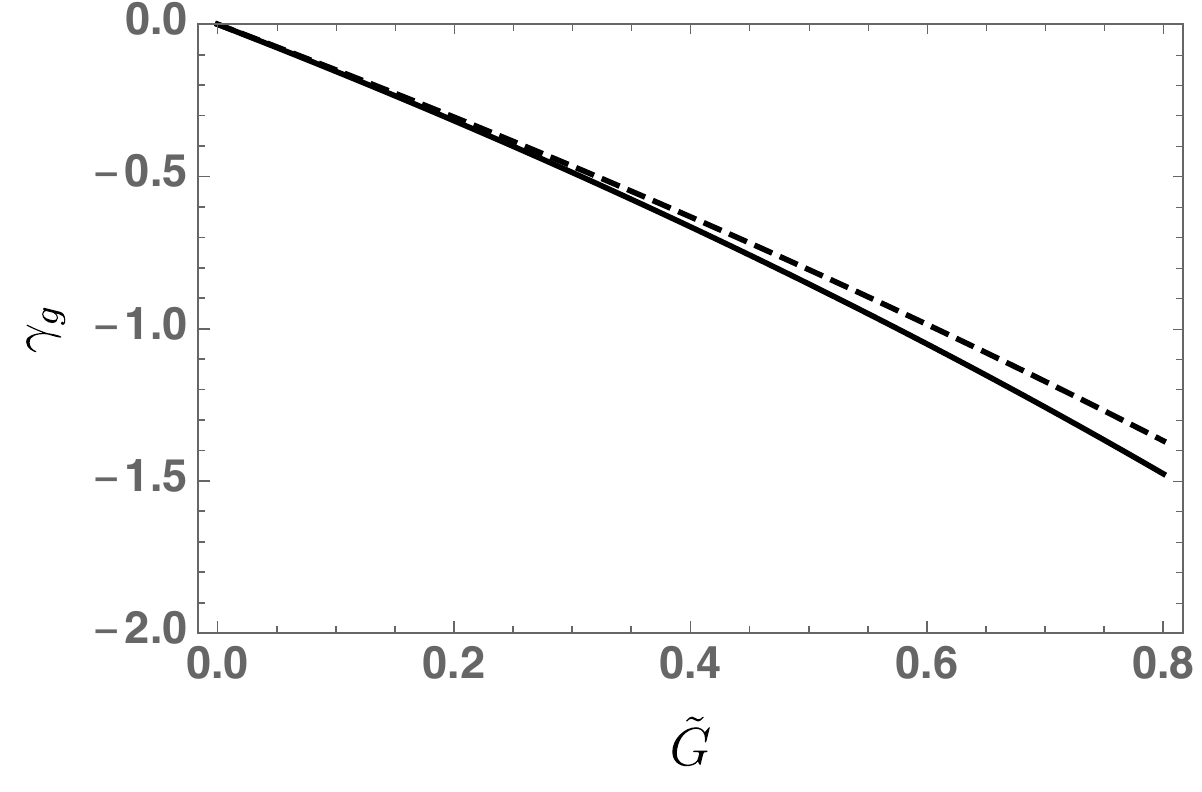}
\caption{The gamma function $\gamma_g$ in the Einstein--Hilbert approximation (dashed line) and the order $\partial^4$ approximation (solid line).}
\label{fig:gamma_g}
\end{figure}

\begin{figure}[h]
\includegraphics[scale=.8]{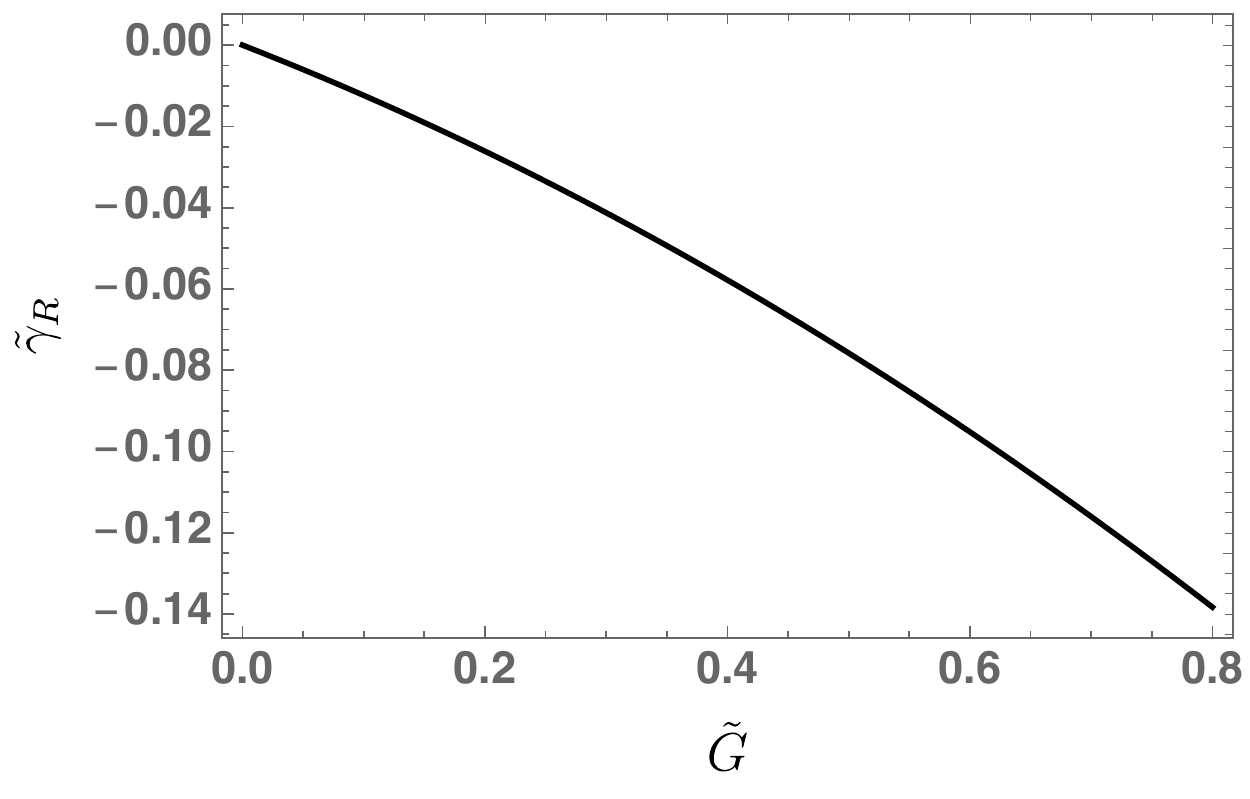}
\caption{The gamma function $\tilde{\gamma}_R$ in the order $\partial^4$ approximation.}
\label{fig:gamma_R}
\end{figure}

\begin{figure}[h]
\includegraphics[scale=.8]{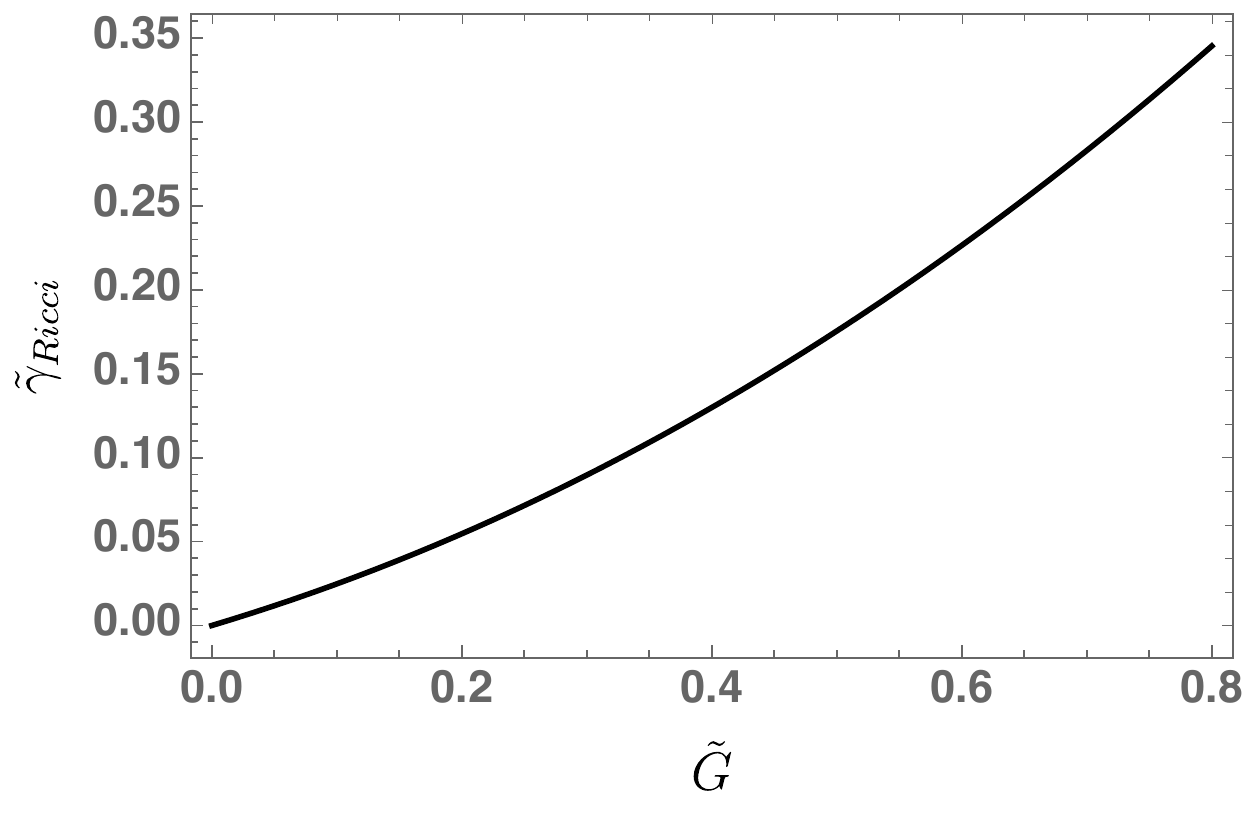}
\caption{The gamma function $\tilde{\gamma}_{Ricci}$ in the order $\partial^4$ approximation.}
\label{fig:gamma_Ricci}
\end{figure}

\begin{figure}[h]
\includegraphics[scale=.8]{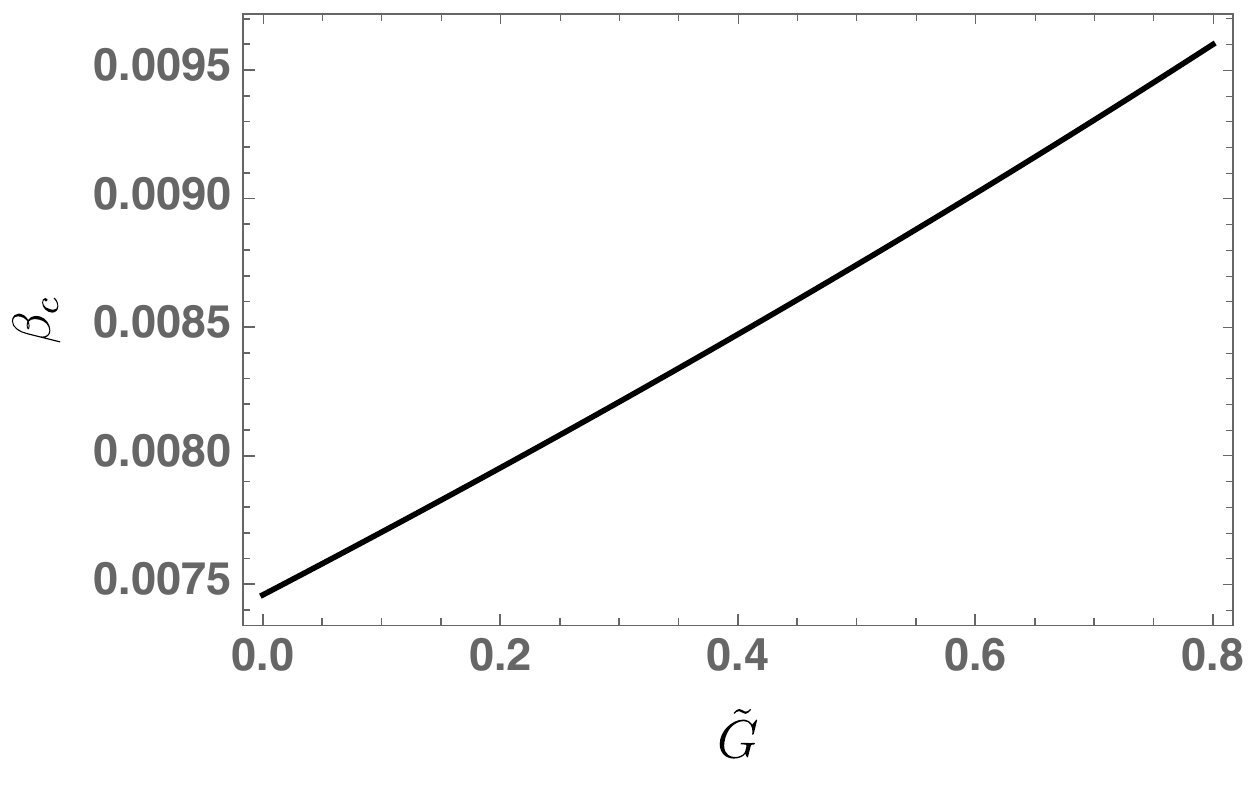}
\caption{The beta function 
 $\beta_c = \partial_t c_k$ in the order $\partial^4$ approximation.}
\label{fig:beta_c}
\end{figure}

The gamma function $\gamma_g(\tilde{G})$, plotted in  Figure~\ref{fig:gamma_g} at orders $\partial^2$ and
$\partial^4$, also appears stable between the two approximations and is approximately linear in weakly coupled phase.
At the Reuter fixed point $\gamma_g$ takes the values   \bea \gamma_g^\star &=& -1.1605 \,\,\,\,\,\, {\rm at }  \,\, O(\partial^2)\,,  \vspace{1cm} \\     \gamma_g^\star &=& -1.1062  \,\,\,\,\,\, {\rm at }  \,\, O(\partial^4)\,.\eea
The stability between the orders can be understood by looking at the gamma functions $\gamma_R(\tilde{G})$ and $\gamma_{Ricci}(\tilde{G})$, which are zero at order $\partial^2$, and remain small at order $\partial^4$ in the region $0<G<\tilde{G}_\star$, as can be seen in  Figures~\ref{fig:gamma_R}~and~\ref{fig:gamma_Ricci}. At the Reuter fixed point $\gamma_R$ and  $\gamma_{Ricci}$  take the values   \bea \gamma_R^\star &=& -0.10079 \,\,\,\,\,\, {\rm at }  \,\, O(\partial^4)\,,  \vspace{1cm} \\     \gamma_{Ricci}^\star &=& 0.24150 \,\,\,\,\,\,\,\, \,\,{\rm at }  \,\, O(\partial^4)\,.\eea Thus, we observe a remarkable stability as the order of the approximation is increased.
At order $\partial^4$ we also find the beta function of $c_k$ which is plotted in Figure~\ref{fig:beta_c}.

 Let us stress that the values of the gamma functions are not universal quantities and will depend on the RG scheme.
 We note that at $\tilde{G}\approx 3$ the beta functions $\beta_{\tilde{G}}( \tilde{G})$ calculated at orders $\partial^2$ and
$\partial^4$ begin to differ substantially. This indicates that the derivative expansion may not converge in the strong coupling phase $\tilde{G} > \tilde{G}_{\star}$. However, since we undoubtedly live in a weakly coupled phase, this should have few phenomenological consequences.

Finally, we note that at the Reuter fixed point the redundant operators~\eqref{Redundant_BF} are given by
\bea \label{RedO2}
\mathcal{T}[g_{\mu\nu}] =  \int \d^dx \sqrt{\det g}  \left( -0.0079673\, k^4 - 0.028948 k^2 \,R\right)\,,
\eea
at order $\partial^2$,
and by
\bea \label{RedO4}
\mathcal{T}[g_{\mu\nu}] &&=  \int \d^dx \sqrt{\det g}  \left(  -0.0079428\, k^4 - 0.030241 k^2 \,R \right.  \\  && \hspace{3cm} \left. - 0.0028418  \, R^2 + 0.0048354\, R_{\mu\nu}R^{\mu\nu} - 0.00072986\, E \right) \,, \nonumber   \\ [5pt]
\mathcal{T}[R g_{\mu\nu}] &&=  \int \d^dx \sqrt{\det g}  \left(  -0.0016664\, k^6 -0.0073873 k^4\,R - 0.029686\,k^2\, R^2 \right) \,, \\ [5pt]
\mathcal{T}[R_{\mu\nu}] &&=  \int \d^dx \sqrt{\det g}  \left( -0.0016664\, k^6 -0.00055327 k^4 \, R \right.    \\  && \hspace{2.5cm} \left. - 0.016115\,k^2\, R^2 + 0.033644\,k^2\, R_{\mu\nu}R^{\mu\nu}  +0.00041544\,k^2\, E\right) \nonumber\,,
\eea
at order $\partial^4$.
It is straightforward to show that these operators~\eqref{RedO2} and~\eqref{RedO4} are linearly independent of the terms in the Reuter fixed point action and form a complete basis at orders  $\partial^2$ and  $\partial^4$, respectively. This confirms that the RG conditions which we choose to fix the inessential couplings at the GFP continue to fix the values of the inessential couplings at the interacting Reuter fixed point.
\vspace{15pt}

\vspace{-9pt}
\vspace{-9pt}
\vspace{-9pt}

\section{Discussion and Outlook}
\label{sec:conclusion}
We have investigated the non-perturbative renormalisability of gravity~\cite{Weinberg} taking care to disregard the running of inessential couplings for the first time. The consequences of doing so are profound: not only are calculations much simpler in the minimal essential scheme, but we also reveal that only Newton's constant is  essential and relevant in our approximation.

 Although this conclusion could change by including  higher-order terms, this seems unlikely since all higher-order terms are canonically irrelevant and, thus, the quantum correction to their scaling dimensions would have to be large.  Additionally, the stability of the fixed point going from order $\partial^2$ to order $\partial^4$ indicates that our approximations do not miss another relevant coupling. Moreover, the Goroff--Sagnotti term, which is the only $\partial^6$ term that is independent of the Ricci curvature, has been found to be irrelevant at the Reuter fixed point~\cite{Gies:2016con}.   As a result we expect that the qualitative picture obtained here at order $s=4$ will not change as we go to higher orders. Ultimately, this can be confirmed by systematically increasing the order of the derivative expansion. This program will be technically simpler within the minimal essential scheme since there will be less terms in the EAA than in the standard approach~\cite{Knorr:2021slg}, which does not remove redundant operators. Furthermore,  it has been argued that additional poles in the propagator can prevent the convergence of the derivative expansion in quantum gravity~\cite{Knorr:2021niv}. However, in the minimal essential scheme we can avoid such poles and thus we expect to see convergence of the derivative expansion as is observed for scalar field theories~\cite{Balog:2019rrg}.  
 
Apart from strengthening the evidence in favour of the existence of the Reuter fixed point, we can also give a straightforward argument in favour of the theory being unitary, since the terms that contain four derivatives are redundant.
 This property will be true of all higher derivatives if the fixed point can be found in the minimal essential scheme, which assumes their absence from the beginning. Consequently, the minimal  essential scheme provides a frame work to address some of the open problems for the asymptotic safety program~\cite{Donoghue:2019clr,Bonanno:2020bil} which concern the form of the propagator.
  We should stress that 
by using the  minimal essential scheme we can dictate which physical degrees of freedom we are attempting to renormalise, and, thus, ensure that we are dealing with a unitary theory, rather than searching in a space of theories littered with non-unitary ones. In calculations that retain terms outside of those in the minimal essential scheme, we expect to find many fixed points which lie in different universality classes. Indeed, studies that include many powers of the Riemann curvature have found fixed points with as many as four relevant directions~\cite{Kluth:2020bdv}. 

Perhaps most profoundly, we have identified the vacuum energy as inessential coupling which agrees with other arguments~\cite{Hamber:2013rb}.
The fact that it is true at the GFP makes this a property of perturbative quantum gravity.
One can show that the contributions proportional to $w_1$ and $w_2$ in the linearised beta function~\eqref{linear_beta_rho} come from the terms proportional to $\Psi_k$ in the RHS of the flow Equation~\eqref{dimensionfull_flow} and terms proportional to $\tilde{\rho}_{\rm GFP}$ from the LHS of the flow equation. Reinstating powers of Planck's constant $\hbar$, one sees that both contributions vanish when $\hbar =0$.
This means that the inessential nature of the vacuum energy is a quantum effect. Indeed, in a scheme where $\tilde{\rho}_{\rm GFP}= 0$ the classical term in the redundant operator would vanish at the GFP, but the contributions due to $\Psi_k$ in the RHS of the flow equation mean $\rho_k$ is anyway inessential.
The elementary understanding of this effect is that a rescaling of field $\hat{g} \to \Omega \hat{g}$ produces an infinite factor $\sim \prod_x \Omega$ in the functional measure which when regularised will renormalise the vacuum energy~\cite{Falls:2017cze}. In the flow equation for the EAA, this manifests in the term proportional to $\gamma_g$ in the RHS of the flow equation. Thus, simply by renormalising the quantum metric field, we can adjust the renormalisation of the vacuum energy. Since, in the universality class we have investigated the vacuum, energy is inessential both at the GFP and the Reuter fixed point, no physical meaning can be attributed to its flow. However, there can be other universality classes, both for pure gravity and for gravity coupled to matter, where the cosmological constant is essential and its flow has physical consequences~\cite{Adeifeoba:2018ydh,Bonanno:2019ilz,Platania:2020lqb}.

Since there is only one relevant essential coupling at the Reuter fixed point, it would appear that the vanishing of the cosmological constant in Planck units  $\tau_0 = \Lambda_k G_k|_{k=0}$ at $k=0$ must be a prediction of the Reuter fixed point. Thus, if a different scheme would find a non-vanishing $\tau_0$ it would be a contradiction that could only be explained as an artefact of an approximation. To investigate this, one can refrain from fixing the renormalisation of $\tilde{\rho}$, as in the minimal essential scheme, but instead only assume $\gamma_g$ vanishes at $\tilde{G}=0$. Then,  expanding $\partial_t \tau_k$ around $\tilde{G}=0$ while keeping $\tau_k$, one finds at order $\partial^2$ that
\beq
\partial_t \tau_k=   -\frac{14 \tilde{G} \tau_k }{3 \pi }  + O(G^2)\,,
\eeq
which implies that $\tau_0$ could take a non-zero value.
 Studying the full beta functions with $\gamma_g=0$ one finds trajectories leaving the Reuter fixed point and ending at any value $\tau_0 <0$, contradicting the minimal essential scheme.
 However, going to order  $\partial^4$ one finds instead that 
\beq
\partial_t \tau_k=   -\frac{328 \tau_k^2}{3 (20 \pi -7 \tau_k )}  + O(G^2)\,,
\eeq
which only vanishes at $\tau_k=0$ and thus no contradiction with the minimal essential scheme can occur. Thus, the vanishing of the observable $\tau_0$ appears to be a robust prediction of the Reuter fixed point.

  Here we have only treated pure gravity and thus to properly address the cosmological constant problem we should understand the situation when matter is coupled to gravity~\cite{Eichhorn:2018yfc}.  Indeed, arguably there was never a cosmological constant problem in pure gravity since if we adopt dimensional regularisation only terms proportional to $\rho_k$ would renormalise $\rho_k$ and we can simply set $\rho_k=0$. What will remain true even in the presence of matter is that there is an inessential coupling related to a rescaling of the spacetime metric. This might shed new light on the cosmological constant problem~\cite{Weinberg:1988cp}.

This work can be extended in several directions. A crucial test is to make sure that the qualitative  picture is stable when the form of the cut-off function is modified. Moreover, to obtain the best numerical estimate of the critical exponent $\uptheta$, the principle of minimal sensitivity (PMS) can be applied by studying the dependence of $\uptheta$ on unphysical parameters, such as those which enter a class of cut-off functions or the values of inessential couplings, such as the vacuum energy. The PMS selects the value of $\uptheta$ where this dependence is minimal (for a recent application of the PMS to the critical exponents of the Ising model see~\cite{Balog:2019rrg}). 
Furthermore, the dependence on the parameterisation of the metric tensor and the choice of gauge~\cite{Gies:2015tca,Ohta:2016npm,Ohta:2016jvw,DeBrito:2018hur} can also be investigated within the minimal essential scheme. In the background field approximation, we neglect the running of these parameters, while a proper treatment of these parameters should identify them with inessential couplings since they cannot enter expressions for observables. Thus, going beyond the background field approximation, the minimal essential scheme should include extra gamma functions in order to impose renormalisation conditions for each unphysical parameter. As an alternative, one can use diffeomorphism and parameterisation invariant exact renormalisation equations, such as those based on the geometrical effective action~\cite{Pawlowski:2003sk,Donkin:2012ud} or the background independent exact renormalisation group~\cite{Falls:2020tmj}.

The finding that there appears to be only one relevant essential coupling in QEG is an encouraging sign for attempts to make contact with other methods which can be used to investigate asymptotic safety. In particular, it would be very interesting if perturbative methods based on expansions around two dimensions~\cite{Falls:2017cze,Kawai:1989yh,Martini:2021slj} could also calculate the critical exponent $\uptheta$ by performing a two-loop calculation in the minimal essential scheme. Additionally, the value of $\uptheta$ can be computed in lattice and tensor model approaches to quantum gravity~\cite{Hamber:2015jja,Laiho:2016nlp,Loll:2019rdj,Eichhorn:2019hsa}.
\vspace{6pt}

\acknowledgments{We thank R.\,Percacci and R.\,Ben Al{\`i} Zinati for a careful reading of the manuscript and for providing us with useful comments and suggestions.}

\begin{appendix}
\section{Calculation at quartic order in minimal essential scheme}
\label{app:calcpuregravity}
In this appendix we derive the flow equations in minimal essential scheme, i.e. the scheme with renormalisation conditions that fix to zero the coefficients of $\sqrt{\det g}\,R^2$ and $\sqrt{\det g}\,R_{\mu\nu} R^{\mu\nu}$ .
Therefore, in such scheme the ansatz for EAA at quartic order is simply
\beq
  \bar{\Gamma}_k[g] =   \int \d^dx  \sqrt{\det g}\left\{ \frac{\rho_k}{8 \pi}
  - \frac{1}{16 \pi G_k} R  + c_k  E \right\} \,,
\eeq
where $\rho_k = \frac{\Lambda_k}{G_k}$ and $E = R_{\mu \nu \a \b}R^{\mu \nu \a \b}-4 R_{\mu \nu}R^{\mu \nu}+ R^2$ .\\
The RG kernel for the quantum metric is given by \eqref{Psi_expansion}, and so the LHS of \eqref{dimensionfull_flow} is equal to
\bea
&&\int \d^dx \sqrt{\det g}
\Big\{ \left( \partial_t +\frac{d}{2} \gamma_g \right)\frac{\rho_k}{8\pi}
+   \left( -\left( \partial_t +\frac{d-2}{2}\gamma_g\right)\frac{1}{16 \pi G_k}
+(\gamma_{Ricci} +d\gamma_R) \frac{\rho_k}{16 \pi} \right)  R
\\
&&\hspace{2cm}-\frac{1}{32 \pi G_k} (\gamma_{Ricci}+(d-2)\gamma_R)R^2
+\frac{\gamma_{Ricci}}{16 \pi G_k} R_{\mu\nu} R^{\mu \nu}
+\left( \partial_t + \frac{d-4}{2}\gamma_g \right)c_k E\Big\} \,.
\nonumber
\eea
The RHS of \eqref{dimensionfull_flow} contains two traces, one coming from the graviton contribution and one from the ghosts contribution and in the following subsections we calculate them, denoting the gravity trace as $\mathbb{T}_{gg}$ and ghost trace as $\mathbb{T}_{\bar c c}$.

As a final remark, note that in the calculations reported below we neglect terms proportional $c_k$ in the traces: this is justified in $d=4$ since in this case the corresponding invariant is topological and so these contributions in RHS of \eqref{dimensionfull_flow} vanish in $d=4$.
\subsection{Calculation of gravity trace}
\label{app:gravitytrace}
In this subsection we calculate the graviton contribution to the quantum part of the flow equation \eqref{dimensionfull_flow}: in particular, we insert the regulator in such a way that $\Delta \to P_k \equiv \Delta + R_k(\Delta)$, we calculate the Hessian, we expand the argument of the trace to quadratic order in curvature and finally we evaluate the trace using off-diagonal heat kernel techniques \cite{Groh:2011dw}.
We then choose the regulator to be given by
\beq
\mathcal{R}_k^{gg} = K_{gg} R_k(\Delta) \,,
\eeq
where
\beq
K_{gg}^{\mu\nu, \a\b} = \frac{1}{2 \kappa^2_k} \, \sqrt{\det g} \left( g^{\mu \a}g^{\nu \b}+g^{\mu \b}g^{\nu \a}-g^{\mu \nu}g^{\a \b} \right) \,,
\eeq
and the following relation holds
\beq
\partial_t K_{gg} = -\eta_N  K_{gg} \,,
\eeq
with $\eta_N = \partial_t  G_k/G_k$.
The Hessian in the gravity sector is
\beq
\frac{\delta^2 \bar{\Gamma}_k}{\delta g \delta g} +   K_{g g} \cdot \Delta_{\rm gf}  + \mathcal{R}^{gg}_k
= K_{gg} \cdot ( P_k  + U_0  + U_1 ) \,,
\eeq
where
\bea
U_0 &=&  -2 \rho_k G_k  \,,
\\
\left(U_1\right){}^{\mu\nu}{}_{\a\b} &=& \frac{1}{2} R \left( \delta^\mu_\a \delta^\nu_\b+\delta^\mu_\b \delta^\nu_\a-g^{\mu \nu}g_{\a \b} \right) + g^{\mu \nu} R_{\a \b}+R^{\mu \nu} g_{\a \b}
-2\delta^{(\mu}_{(\a}R^{\nu)}_{\b)}-2R^{(\mu}{}_{(\a}{}^{\nu)}{}_{\b)}
\nn
&&-\,\frac{d-4}{d-2} \,g_{\a \b}\,\left( R^{\mu \nu} -\frac{1}{2} R g^{\mu \nu}\right) \,,
\eea
and the indices in the round brackets are symmetrised. 
Then the gravitational trace is given by
\bea
\mathbb{T}_{gg} &=& \frac{1}{2} \Tr \frac{1}{P_k(\Delta) +  U_0 + U_1 }  \cdot \left( ( \partial_t -\eta_N)  R_k(\Delta) + 2  \frac{\delta}{\delta g}  \Psi_k^g \cdot \,R_k(\Delta) \right)
\\
 &=&   \frac{1}{2} \Tr \, \Big\{ \G_k(\Delta)  -  \G_k(\Delta)^2   U_1   +   \G_k(\Delta)^3  U_1^2       \Big\} \Big\{ (  \partial_t -\eta_N)  R_k(\Delta) + 2  ( V^{\mu\nu} \nabla_{(\mu} \nabla_{\nu)}   + W_0 + W_1  ) R_k(\Delta) \Big\} \,,
 \nonumber
\eea
where we have written
\bea
\G_k(\Delta) &=&  \frac{1}{P_k(\Delta) +  U_0} \,,
\\
\frac{\delta}{\delta g}  \Psi_k^g &=& V + W = V^{\mu\nu} \nabla_{(\mu} \nabla_{\nu)}  + W_0 + W_1 \,,
\\
V^{\mu \nu}{}_{\rho \sigma}{}^{\a \b} &=&
\gamma_{Ricci} \left( -\frac{1}{2} \delta^{\mu}_\rho \delta^{\nu}_\sigma g^{\a \b}
+ \delta^{\mu}_{(\rho} g^{\nu (\a} \delta_{\sigma)}^{\b)}
- \frac{1}{2}g^{\mu \nu}\delta_{(\rho}^{(\a}\delta_{\sigma)}^{\b)}  \right)
+\gamma_R \, g_{\rho \sigma} \left(
g^{\mu \a} g^{ \b \nu}-g^{\mu \nu} g^{\a \b}\right) \,,
\\
(W_0){}_{\rho \sigma}{}^{\a \b} &=& \gamma_g \delta_{(\rho}^{(\a}\delta_{\sigma)}^{\b)} \,,
\\
(W_1){}_{\rho \sigma}{}^{\a \b} &=&   \frac{1}{2} \gamma_{Ricci}\left( \delta_{(\rho}^{(\a}R_{\sigma)}^{\b)}-R^{(\a}{}_{(\rho}{}^{\b)}{}_{\sigma)} \right)
 + \gamma_R \left( R\, \delta_{(\rho}^{(\a}\delta_{\sigma)}^{\b)} -g_{\rho \sigma} R^{\a \b} \right) \,.
\eea
Defining $\dot R_k := (  \partial_t-\eta_N )  R_k(\Delta)$ , $\mathbb{T}_{gg}$ is composed by nine traces, which read
\bea
&&\Big\{\frac{1}{2} \Tr \,    \G_k  \dot  R_k(\Delta) \,;\,
-\frac{1}{2} \Tr \,    \G_k^2  U_1  \dot  R_k(\Delta) \,;\,
\frac{1}{2} \Tr \,    \G_k^3  U_1^2 \dot R_k(\Delta) \,;\,
\nn
&& \hspace{0.5cm}\Tr \,    \G_k  W R_k(\Delta)  \,;\,
 -  \Tr \,    \G_k^2 U_1  W R_k(\Delta) \,;\,
 \Tr \,    \G_k^3 U_1^2  W R_k(\Delta) \,;\,
 \nn
 && \hspace{0.5cm}
  \Tr \,    \G_k  V^{\mu\nu} \nabla_{(\mu} \nabla_{\nu)} R_k(\Delta) \,;\,
  - \Tr \,    \G_k^2 U_1  V^{\mu\nu} \nabla_{(\mu} \nabla_{\nu)} R_k(\Delta) \,;\,
\Tr \,    \G_k^3 U_1^2  V^{\mu\nu} \nabla_{(\mu} \nabla_{\nu)} R_k(\Delta) \,
\Big\}\,.
\nonumber
\eea
Defining
\beq
Q_{n}[W(\Delta)] := \frac{1}{\Gamma(n)} \int_0^\infty \d z
z^{n-1} W(z) \,,
\eeq
below we report the evaluation of these traces
\begin{equation}
\begin{split}
\bullet \hspace{0.3cm}&
\frac{1}{2} \Tr \,    \G_k  \dot  R_k(\Delta)   =  \frac{1}{(4 \pi)^{d/2}}  \frac{1}{2} \sum_n Q_{d/2-n}\left[ \G_k \dot{R}_k   \right] \tr A_n \\
&=  \frac{1}{2(4 \pi)^{d/2}} \int \d^dx \sqrt{\det g} \left\{   \frac{d(d+1)}{2}  Q_{d/2}\left[ \G_k \dot{R}_k  \right]
+   \frac{d(d+1)}{12}  \, R \, Q_{d/2-1}\left[ \G_k \dot{R}_k \right] \right. \nn
& \left. \hspace{2cm}+   \frac{1}{180}  \,
\left( \frac{d(d+1)}{2} \left( \frac{5}{2}R^2
- R_{\mu \nu} R^{\mu \nu} \right)
+\frac{d^2-29d-60}{2} R_{\mu \nu \a \b} R^{\mu \nu \a \b}\right)
Q_{d/2-2}\left[ \G_k \dot{R}_k  \right] \right\} \,,
\nonumber\\
\bullet \hspace{0.3cm}&
-\frac{1}{2} \Tr \,    \G_k^2  U_1  \dot  R_k(\Delta)   =   -   \frac{1}{(4 \pi)^{d/2}}  \frac{1}{2} \sum_n Q_{d/2-n}\left[ \G_k^2 \dot{R}_k \right] \tr U_1 A_n
\hspace{5cm}\\
&= -   \frac{1}{2(4 \pi)^{d/2}} \int \d^dx \sqrt{\det g}\left \{  \frac{d(d-1)}{2} R \,Q_{d/2}\left[ \G_k^2 \dot{R}_k \right]
+  \frac{d(d-1)}{12}\, R^2 \,Q_{d/2-1}\left[ \G_k^2 \dot{R}_k \right]\right\}\,,
\nonumber\\
\bullet \hspace{0.3cm}&
\frac{1}{2} \Tr \,    \G_k^3  U_1^2 \dot R_k(\Delta)   =
\frac{1}{(4 \pi)^{d/2}}    \frac{1}{2} \sum_n Q_{d/2-n}\left[ \G_k^3 \dot{R}_k  \right] \tr U_1^2 A_n
\label{trace3}\\
&= \frac{1}{2(4 \pi)^{d/2}}
\int \d^dx \sqrt{\det g}
\left\{ \frac{d^3-5d^2+8d+4}{2(d-2)} R^2
 +\frac{d^2-8d+4}{d-2}  R_{\mu \nu} R^{\mu \nu}
+3 R_{\mu \nu \a \b} R^{\mu \nu \a \b}\right\}Q_{d/2}\left[ \G_k^3 \dot{R}_k \right] \,,
\nonumber\\
\bullet \hspace{0.3cm}&
\Tr \,    \G_k  W R_k(\Delta) =  \frac{1}{(4 \pi)^{d/2}}   \sum_n Q_{d/2-n}\left[ \G_k R_k  \right] \tr W A_n \\
&=
 \frac{\gamma_g}{(4 \pi)^{d/2}} \int \d^dx \sqrt{\det g} \left\{  \frac{d(d+1)}{2}  Q_{d/2}\left[ \G_k R_k  \right]
+   \frac{d(d+1)}{12}  R\,Q_{d/2-1}\left[ \G_k R_k \right]  \right. \nn
& \left. \hspace{2cm}+   \frac{1}{180}
\left( \frac{d(d+1)}{2} \left( \frac{5}{2}R^2
- R_{\mu \nu} R^{\mu \nu} \right)
+\frac{d^2-29d-60}{2} R_{\mu \nu \a \b} R^{\mu \nu \a \b}\right)
Q_{d/2-2}\left[ \G_k R_k  \right] \right\} \nn
& \hspace{0.5cm}+  \frac{1}{(4 \pi)^{d/2}} \int \d^dx\sqrt{\det g}
\left(  \gamma_{Ricci}+2(d-1)\gamma_R  \right)\frac{d+2}{4}
\left\{
  R \,Q_{d/2}\left[ \G_k R_k  \right]
+
\frac{1}{6}
 R^2 \, Q_{d/2-1}\left[ \G_k R_k \right]\right\} \,,
\nonumber
\end{split}
\end{equation}
\begin{equation}
\begin{split}
\bullet \hspace{0.3cm}&
-  \Tr \,    \G_k^2 U_1  W R_k(\Delta)   = -  \frac{1}{(4 \pi)^{d/2}}   \sum_n Q_{d/2-n}\left[ \G_k^2 R_k  \right] \tr U_1 W A_n\\
&= -   \frac{\gamma_g}{(4 \pi)^{d/2}} \int \d^dx \sqrt{\det g}\left \{  \frac{d(d-1)}{2} R\,Q_{d/2}\left[ \G_k^2 R_k \right]
+  \frac{d(d-1)}{12} R^2 \,Q_{d/2-1}\left[ \G_k^2 R_k \right]\right\}
\nn
& \hspace{0.5cm}-   \frac{1}{(4 \pi)^{d/2}} \int \d^dx \sqrt{\det g}\left \{
 \left( \frac{(d+1)}{4}\gamma_{Ricci} +  \left(\frac{d(d-1)}{2}-\frac{d-4}{d-2}\right)\gamma_R \right)R^2 \right.
\nn
&  \left.\hspace{2cm} + \left( -\frac{(d+2)}{4}\gamma_{Ricci} + 2\left(\frac{d-4}{d-2}\right)\gamma_R  \right) R_{\mu \nu} R^{\mu \nu}+\frac{3\gamma_{Ricci}}{4}R_{\mu \nu \a \b} R^{\mu \nu \a \b} \right\}Q_{d/2}\left[ \G_k^2 R_k \right]  \,,
\nonumber\\
\bullet \hspace{0.3cm}&
\Tr \,    \G_k^3 U_1^2  W R_k(\Delta)   =   \frac{1}{(4 \pi)^{d/2}}   \sum_n Q_{d/2-n}\left[ \G_k^3 R_k  \right] \tr U_1^2 W A_n \\
&=\frac{\gamma_g}{(4 \pi)^{d/2}}
\int \d^dx \sqrt{\det g}
\left\{ \frac{d^3-5d^2+8d+4}{2(d-2)} R^2
 +\frac{d^2-8d+4}{d-2}  R_{\mu \nu} R^{\mu \nu}
+3 R_{\mu \nu \a \b} R^{\mu \nu \a \b}\right\}
Q_{d/2}\left[ \G_k^3 R_k \right] \,,
\nonumber
\end{split}
\end{equation}
\begin{equation}
\begin{split}
\bullet \hspace{0.3cm}&
\Tr \,    \G_k  V^{\mu\nu} \nabla_{(\mu} \nabla_{\nu)} R_k(\Delta)   = \frac{1}{(4 \pi)^{d/2}}   \sum_n Q_{d/2+1-n}\left[ \G_k R_k  \right] \left( -\frac{1}{2} \tr V^{\mu}\,_\mu A_n   +  V^{\mu\nu} A_{n-1|\mu\nu}    \right)
\hspace{1cm}\\
&=
-\frac{1}{2(4 \pi)^{d/2}}  \int \d^dx \sqrt{\det g}\left\{  d(1-d)\left(\frac{d}{4}\gamma_{Ricci} +\gamma_R  \right)
\left( Q_{d/2+1}\left[ \G_k R_k  \right] +\frac{R}{6} Q_{d/2}\left[ \G_k R_k  \right] \right)
\right.\nn
&\hspace{2cm}\left.+
\frac{1}{180}\left\{
d(1-d)\left(\frac{d}{4}\gamma_{Ricci} +\gamma_R  \right) \left(  \frac{5}{2}R^2 - R_{\mu \nu} R^{\mu \nu} \right)
\right.\right.\nn
& \left.\left.\hspace{2cm}
+\left( \frac{-d^3+31d^2-120}{4}\gamma_{Ricci}+ d(1-d) \gamma_R \right) R_{\mu \nu \a \b} R^{\mu \nu \a \b}   \right\}  Q_{d/2-1}\left[ \G_k R_k  \right] \right\}
\nn
& \hspace{0.5cm}
+\frac{1}{(4 \pi)^{d/2}}\int \d^dx \sqrt{\det g} \left\{ \,
(1-d)\left(  \frac{d}{4}\gamma_{Ricci} + \gamma_R  \right) \frac{R}{6}
\,Q_{d/2}\left[ \G_k R_k  \right]
\right.\nn
&\left.\hspace{2cm}
+  \left\{
\frac{\gamma_{Ricci}}{24} \left( -(d+4) R_{\mu \nu} R^{\mu \nu}
+ \frac{3d}{2} R_{\mu \nu \a \b} R^{\mu \nu \a \b} \right) \right.\right.\nn
& \hspace{2cm} \left.\left.
+ \frac{(1-d)}{90}\left(  \frac{d}{4}\gamma_{Ricci} + \gamma_R  \right) \left( \frac{5}{2}R^2 -  R_{\mu \nu} R^{\mu \nu}+ R_{\mu \nu \a \b} R^{\mu \nu \a \b}\right)\right\}Q_{d/2-1}\left[ \G_k R_k  \right]
\right\} \,,
\nonumber\\
\bullet \hspace{0.3cm}&
- \Tr \,    \G_k^2 U_1  V^{\mu\nu} \nabla_{(\mu} \nabla_{\nu)} R_k(\Delta)   =
-  \frac{1}{(4 \pi)^{d/2}}   \sum_n  \left(  -\frac{1}{2} \tr U_1 V^{\mu}\,_\mu A_n   +  U_1 V^{\mu\nu} A_{n-1|\mu\nu}    \right)Q_{d/2+1-n}\left[ \G_k^2 R_k  \right]
\hspace{2cm}\\
&=
\frac{1}{2(4 \pi)^{d/2}}
\int \d^dx \sqrt{\det g}
\left( \frac{-d^3+3d^2-4d+8}{4}\gamma_{Ricci} -( d-4)(d-1) \gamma_R \right)
\nn
& \hspace{6cm} \times
\left(R
\,Q_{d/2+1}\left[ \G_k^2 R_k  \right]
+ \frac{R^2}{6}
\,Q_{d/2}\left[ \G_k^2 R_k  \right] \right)
\nn
& \hspace{0.5cm}
-  \frac{1}{(4 \pi)^{d/2}}
\int \d^dx \sqrt{\det g}
\left\{ -\frac{1}{24(d-2)}\left( (d^3-5d^2+6d+4)\gamma_{Ricci}
+4(d-3)(d-4)\gamma_R \right) R^2
\right.\nn
& \left.\hspace{5cm}-\frac{(d-4)}{6(d-2)} \left( (d-1)\gamma_{Ricci} + 2\gamma_R \right) R_{\mu \nu}R^{\mu \nu}
\right\}Q_{d/2}\left[ \G_k^2 R_k  \right] \,,
\nonumber
\end{split}
\end{equation}
and finally
\begin{equation}
\begin{split}
\bullet \hspace{0.3cm}&
\Tr \,    \G_k^3 U_1^2  V^{\mu\nu} \nabla_{(\mu} \nabla_{\nu)} R_k(\Delta)   =  \frac{1}{(4 \pi)^{d/2}}   \sum_n Q_{d/2+1-n}\left[ \G_k^3 R_k  \right] \left(  -\frac{1}{2} \tr U_1^2 V^{\mu}\,_\mu A_n   +  U_1^2 V^{\mu\nu} A_{n-1|\mu\nu}    \right)
\\
&=
- \frac{1}{2(4 \pi)^{d/2}}
\int \d^dx \sqrt{\det g}
\left\{ -\frac{1}{4(d-2)} \left[ (d^4-7d^3+20d^2-28d+24)\gamma_{Ricci}+4(d-1)(d-4)^2\gamma_R \right] R^2
\right.\nn
&
\left. \hspace{4cm}
+ \frac{1}{2(d-2)} \left[ -(d^3-12d^2+36d-40)\gamma_{Ricci}+4(d-1)(d-4)^2\gamma_R \right]  R_{\mu \nu} R^{\mu \nu}
\right.\nn
&
\left.\hspace{4cm}+3 \left( 1-\frac{d}{2} \right) \gamma_{Ricci} R_{\mu \nu \a \b} R^{\mu \nu \a \b}
\right\}Q_{d/2+1}\left[ \G_k^3 R_k  \right] \,.
\nonumber
\end{split}
\end{equation}
\subsection{Calculation of ghost trace}
\label{app:ghosttrace}
In this subsection we calculate the ghosts contribution to the quantum part of the flow equation \eqref{dimensionfull_flow}: like in the previous subsection, we insert the regulator in such a way that $\Delta \to P_k \equiv \Delta + R_k(\Delta)$, we calculate the Hessian, we expand the argument of the trace to quadratic order in curvature and finally we evaluate the trace.
We then choose the regulator to be given by
\beq
\mathcal{R}_k^{\bar c c} = K_{\bar c c} R_k(\Delta) \,,
\eeq
where
\beq
K^{\mu\nu}_{\bar{c} c} =\frac{\sqrt{2} }{\kappa_k} \sqrt{\det g} g_{\mu \nu} \,,
\eeq
and the following relation holds
\beq
\partial_t K_{\bar c c} = -\frac{\eta_N}{2}  K_{\bar c c} \,.
\eeq
Since the Hessian in the ghost sector is
\begin{align}
K_{\bar{c} c} \cdot \Delta_{\rm gh}  + \mathcal{R}_k^{\bar{c}c } =
K_{\bar c c} \cdot \left( P_k  - Ricci \right) \,,
\end{align}
the ghost trace is given by
\begin{equation}
\begin{split}
& \mathbb{T}_{\bar c c}=- \Tr \left( \frac{1}{P_k}  +   Ricci \frac{1}{P_k^2}  + Ricci^2 \frac{1}{P_k^3}\right)   (\partial_t R_k - \frac{1}{2} \eta_N R_k) =
\\
&=-\frac{1}{(4\pi)^{d/2}}\int \d^dx \sqrt{\det g} \left\{ d\,Q_{d/2}\left[ \frac{(\partial_t R_k - \frac{1}{2} \eta_N R_k)}{P_k}  \right]
+\frac{d}{6}  R\, Q_{d/2-1}\left[\frac{(\partial_t R_k - \frac{1}{2} \eta_N R_k)}{P_k}  \right]
\right.\nn
&\left. \hspace{1cm}
+  \frac{1}{180} \left( \frac{5d}{2}R^2 -d R_{\mu \nu}R^{\mu \nu} +(d-15) R_{\mu \nu \a \b}R^{\mu \nu \a \b} \right)Q_{d/2-2}\left[\frac{(\partial_t R_k - \frac{1}{2} \eta_N R_k)}{P_k}  \right]
\right.\nn
&\left.\hspace{1cm}
+   R\,Q_{d/2}\left[\frac{\left(\partial_t - \frac{1}{2} \eta_N \right)R_k}{P_k^2}  \right]
+  \frac{1}{6} R^2\,Q_{d/2-1}\left[\frac{\left(\partial_t - \frac{1}{2} \eta_N \right)R_k}{P_k^2}  \right]
+   R_{\mu \nu}R^{\mu \nu} \,Q_{d/2}\left[\frac{\left(\partial_t - \frac{1}{2} \eta_N \right)R_k}{P_k^3}  \right]
\right\} \,.
\nonumber
\end{split}
\end{equation}
\subsection{Beta and gamma functions}
\label{app:graflowequations}
In this subsection we put all the contributions inside the flow equation together and we write down the beta functions for $\rho_k$, $G_k$ and $c_k$ and the equations for the gamma functions $\gamma_{Ricci}$ and $\gamma_R$.
In order to express everything in the curvature basis $\left( R^2 , R_{\mu\nu}R^{\mu\nu} , E \right)$, we have expressed the Riemann tensor square as $ R_{\mu \nu \a \b}R^{\mu \nu \a \b}=E+4 R_{\mu \nu}R^{\mu \nu}- R^2 $ in the equations contained in \ref{app:gravitytrace} and \ref{app:ghosttrace}.
From the coefficient of $\sqrt{\det g}$, we can find the beta function of $\rho_k$ by solving
\begin{align}
\label{eq:flowrho}
\left( \partial_t +\frac{d}{2} \gamma_g \right)\frac{\rho_k}{8\pi}
&=
\frac{1}{(4 \pi)^{d/2}}
\left\{\frac{d(d+1)}{2}
\left( \frac{1}{2}Q_{d/2}\left[ \G_k \dot{R}_k  \right] +\gamma_g Q_{d/2}\left[ \G_k R_k  \right] \right)
\right.\\
&\left.\hspace{2cm}
+\frac{d(d-1)}{2} \left(\frac{d}{4}\gamma_{Ricci} +\gamma_R  \right) Q_{d/2+1}\left[ \G_k R_k  \right]
\right.\nn
&\left.\hspace{2cm}
-d\, Q_{d/2}\left[ \frac{\left(\partial_t - \frac{1}{2} \eta_N \right)R_k}{P_k}  \right]  \right\}\,.
\nonumber
\end{align}
Note that \eq{eq:flowrho} can be also understood as an equation for $\gamma_g$: in fact, it is possible to fix the value of $\tilde \rho_k$ tuning $\gamma_g$. As we discussed in section~\ref{sec:vaceniness}, this procedure corresponds to impose a renormalisation condition that fix the value of the vacuum energy. \\
From the coefficient of $\sqrt{\det g}\, R$, we can find the beta function of $G_k$
\begin{align}
\label{eq:flowG}
&-\left( \partial_t +\frac{d-2}{2}\gamma_g\right)\frac{1}{16 \pi G_k}
+(\gamma_{Ricci} +d\gamma_R) \frac{\rho_k}{16 \pi}
=\\
&\hspace{2cm}=\frac{1}{(4 \pi)^{d/2}}
\left\{\frac{d(d+1)}{12} \left( \frac{1}{2}Q_{d/2-1}\left[ \G_k \dot{R}_k  \right]  +\gamma_g Q_{d/2-1}\left[ \G_k R_k  \right]  \right)
\right.\nn
&\left.\hspace{3cm}
-  \frac{d(d-1)}{2} \left( \frac{1}{2}Q_{d/2}\left[ \G_k^2 \dot{R}_k \right] +\gamma_g Q_{d/2}\left[ \G_k^2 R_k \right] \right)
\right.\nn
&\left.\hspace{3cm}
+
\frac{1}{48} \left( \left(d^3-3 d^2+14 d+24\right)\gamma_{Ricci}+4  \left(7 d^2+3 d-10\right) \gamma_R \right)
Q_{d/2}\left[ \G_k R_k  \right]
\right.\nn
&\left.\hspace{3cm}
+\frac{1}{2}
\left( \frac{-d^3+3d^2-4d+8}{4}\gamma_{Ricci} -( d-4)(d-1) \gamma_R \right)
Q_{d/2+1}\left[ \G_k^2 R_k  \right]
\right.\nn
& \left.\hspace{3cm}
- \frac{d}{6} Q_{d/2-1}\left[\frac{\left(\partial_t - \frac{1}{2} \eta_N \right)R_k}{P_k}  \right]
- Q_{d/2}\left[\frac{\left(\partial_t - \frac{1}{2} \eta_N \right)R_k}{P_k^2}  \right]  \right\}\,.
\nonumber
\end{align}
The coefficient of $\sqrt{\det g}\, R^2$ is
\begin{align}
\label{eq:gammarsca}
&-\frac{1}{32 \pi G_k}(\gamma_{Ricci}+(d-2)\gamma_R)
=\nn
&\hspace{2cm}=
\frac{1}{(4 \pi)^{d/2}} \left\{
\frac{d^2+21 d+40}{240}
\left( \frac{1}{2}Q_{d/2-2}\left[ \G_k \dot{R}_k  \right]  +\gamma_g Q_{d/2-2}\left[ \G_k R_k  \right]  \right)
\right.\\
&\left. \hspace{3cm}
-   \frac{d(d-1)}{12}\left( \frac{1}{2}Q_{d/2-1}\left[ \G_k^2 \dot{R}_k \right]+\gamma_g Q_{d/2-1}\left[ \G_k^2 R_k \right] \right)
\right.\nn
&\left. \hspace{3cm}
+ \frac{d^3-5 d^2+2 d+16}{2 (d-2)}
\left( \frac{1}{2}Q_{d/2}\left[ \G_k^3 \dot{R}_k \right]+\gamma_g Q_{d/2}\left[ \G_k^3 R_k \right]\right)
\right.\nn
&\left. \hspace{3cm}
+ \frac{1}{960}\left(   (d-1) d (d+16)\gamma_{Ricci}+12  (d-1) (7 d+12)\gamma_R \right)  Q_{d/2-1}\left[ \G_k R_k \right]
\right.\nn
&\left. \hspace{3cm}
-\frac{ \left(d^4-7 d^3+32 d^2-76 d+56\right)\gamma_{Ricci}+4  \left(7 d^3-27 d^2+28 d+16\right)\gamma_R}{48 (d-2)}    Q_{d/2}\left[ \G_k^2 R_k \right]
\right.\nn
&\left. \hspace{3cm}
+ \frac{d \left(d^3-7 d^2+14 d-4\right)\gamma_{Ricci} +4 (d-1) (d-4)^2\,\gamma_R}{8 (d-2)}
Q_{d/2+1}\left[ \G_k^3 R_k  \right]
\right.\nn
& \left. \hspace{3cm}
- \frac{d+10}{120}
Q_{d/2-2}\left[\frac{\left(\partial_t - \frac{1}{2} \eta_N \right)R_k}{P}  \right]
- \frac{1}{6}
Q_{d/2-1}\left[\frac{\left(\partial_t - \frac{1}{2} \eta_N \right)R_k}{P^2}  \right]
\right\} \,,
\nonumber
\end{align}
and the coefficient of $\sqrt{\det g}\, R_{\mu\nu}R^{\mu\nu}$ is
\begin{align}
\label{eq:gammaricci}
\frac{\gamma_{Ricci}}{16 \pi G_k}
&=
\frac{1}{(4 \pi)^{d/2}}
\left\{
\frac{\left(d^2-39 d-80\right)}{120}
\left( \frac{1}{2}Q_{d/2-2}\left[ \G_k \dot{R}_k  \right]+\gamma_g Q_{d/2-2}\left[ \G_k R_k  \right]\right)
\right.\\
& \left.\hspace{1cm}
+
\frac{d^2+4 d-20}{d-2}\left( \frac{1}{2}Q_{d/2}\left[ \G_k^3 \dot{R}_k \right]+\gamma_g Q_{d/2}\left[ \G_k^3 R_k \right]\right)
\right.\nn
& \left.\hspace{1cm}
+
\frac{1}{480} \left(\left(d^3-45 d^2+104 d+80\right)\gamma_{Ricci}+
4  \left(d^2-5 d+4\right)\gamma_R \right)
Q_{d/2-1}\left[ \G_k R_k  \right]
\right.\nn
& \left.\hspace{1cm}
+
 \frac{  \left(5 d^2-46 d+68\right)\gamma_{Ricci}-20  (d-4)\gamma_R}{12 (d-2)}
Q_{d/2}\left[ \G_k^2 R_k \right]
\right.\nn
& \left.\hspace{1cm}
+\frac{ \left(d^3-12 d+8\right)\gamma_{Ricci}-4  (d-4)^2 (d-1)\gamma_R}{4 (d-2)} Q_{d/2+1}\left[ \G_k^3 R_k  \right]
\right.\nn
& \left.\hspace{1cm}
-  \frac{d-20}{60}
Q_{d/2-2}\left[\frac{\left(\partial_t - \frac{1}{2} \eta_N \right)R_k}{P_k}  \right]
- Q_{d/2}\left[\frac{\left(\partial_t - \frac{1}{2} \eta_N \right)R_k}{P^3}  \right]
\right\} \,.
\nonumber
\end{align}
Note that \eqref{eq:gammarsca} and \eqref{eq:gammaricci} are the equations for the gamma functions $\gamma_{Ricci}$ and $\gamma_R$, which are the parameters of the RG kernel that fix to zero the value of the couplings associated to the operators $\sqrt{\det g}\, R^2$ and $\sqrt{\det g}\, R_{\mu\nu}R^{\mu\nu}$.\\
Finally, from the coefficient of $\sqrt{\det g}\, E$ we can find the beta function of $c_k$
\bea
\label{eq:flowc}
\left( \partial_t + \frac{d-4}{2}\gamma_g \right)c_k
&=&
\frac{1}{(4 \pi)^{d/2}}  \left\{ \frac{d^2-29d-60}{360} \left( \frac{1}{2}Q_{d/2-2}\left[ \G_k \dot{R}_k  \right] +\gamma_g Q_{d/2-2}\left[ \G_k R_k  \right]  \right)
\right.\\
&&\left.\hspace{1cm}
+ 3 \left( \frac{1}{2}Q_{d/2}\left[ \G_k^3 \dot{R}_k \right] +\gamma_g Q_{d/2}\left[ \G_k^3 R_k \right] \right)
-  \frac{3\gamma_{Ricci}}{4} Q_{d/2}\left[ \G_k^2 R_k \right]
\right.\nn
&& \left.\hspace{1cm}
+\frac{(d-4) }{1440}
\left(  \left(d^2-31 d-30\right)\gamma_{Ricci}+4 (d-1)\gamma_R \right)
Q_{d/2-1}\left[ \G_k R_k  \right]
\right.\nn
&& \left. \hspace{1cm}
- \frac{3}{2}
\left( 1-\frac{d}{2} \right) \gamma_{Ricci} Q_{d/2+1}\left[ \G_k^3 R_k  \right]
- \frac{(d-15)}{180}
Q_{d/2-2}\left[\frac{\left(\partial_t - \frac{1}{2} \eta_N \right)R_k}{P_k}  \right]
\right\} \,.
\nonumber
\eea
\end{appendix}

\newpage


\end{document}